\documentclass[a4paper,11pt]{article}
\pdfoutput=1 

\usepackage{jcappub} 

\usepackage{lmodern}
\usepackage[T1]{fontenc} 
\usepackage{subfig}

\usepackage{color}
\usepackage[dvipsnames]{xcolor}
\usepackage{hyperref}
\usepackage[normalem]{ulem} 

\usepackage{orcidlink}

\DeclareUnicodeCharacter{0456}{\i}

\title{Testing time-delayed cosmology}

\author[a,1]{C. J. Palpal-latoc,\orcidlink{0000-0002-9171-6252}\note{Corresponding Author}}
\author[b]{Reginald Christian Bernardo,\orcidlink{0000-0001-8589-6851}}
\author[a]{and Ian Vega\orcidlink{0000-0002-0428-048X}}

\affiliation[a]{National Institute of Physics, University of the Philippines,\\ Diliman, Quezon City 1101, Philippines}
\affiliation[b]{Institute of Physics, Academia Sinica,\\ Taipei 11529, Taiwan}

\emailAdd{cpalpallatoc@nip.upd.edu.ph}
\emailAdd{reginaldchristianbernardo@gmail.com}
\emailAdd{ivega@nip.upd.edu.ph}

\abstract{Motivated by the proposed time-delayed cosmology in the primordial inflationary era, we consider the application of the delayed Friedmann equation in the late-time Universe and explore some of its observable consequences. We study the background evolution predicted by the delayed Friedmann equation and determine the growth of Newtonian perturbations in this delayed background. We reveal smoking-gun imprints of time-delayed cosmology that can be traced to derivative discontinuities generic in delay differential equations. We show that a late-time cosmic delay is statistically consistent with Hubble expansion rate and growth data. Based on these observables, we compute a nonzero best estimate for the time delay parameter and find that the Bayesian evidence does not strongly rule out a late-time time delay but warrants the subject further study.}

\keywords{time-delayed cosmology, structure formation, inflation, modified gravity, delay differential equations}

\begin{document}
\maketitle
\flushbottom

\section{Introduction}\label{introduction}
The discovery of the cosmic microwave background stamped the Big Bang model as a canonical theory of cosmic evolution. But despite its success, the Big Bang model is afflicted by several problems. These include the flatness problem, which is the problem of why the Universe on very large scales is approximately spatially flat today and even flatter before \cite{Hawking:2010mca}; the horizon problem, which is the problem of causally disconnected regions in the sky having the same temperature \cite{Rindler:1956yx, Weinberg:1972kfs}; and the monopole problem, which refers to the absence of observed magnetic monopoles \cite{Zeldovich:1978wj}. These problems are not pathologies of the Big Bang per se, but the model does not have the predictive power to solve them. Therefore, the need for an addendum to the Big Bang model is widely supported.  

The mainstream resolution to these cosmic conundrums is inflation \cite{Guth:1980zm, Linde:1981mu}. Inflation posits that the early Universe underwent exponential expansion. This accelerated expansion drove down the initial curvature of spacetime, locked in the uniformity of the Universe, and diluted the density of magnetic monopoles to negligible levels all at once. But despite the elegance of the theory, the usual implementation of inflation via scalar fields called inflatons comes with its own problems \cite{Chowdhury:2019otk, Turok:2002yq}. Inflaton models usually violate energy conditions \cite{Kontou:2020bta, Maleknejad:2012as}, and the fundamental nature of inflatons also remains an open question \cite{Sloan:2018osd, Steinwachs:2019hdr}. These reasons continue to motivate the search for alternative mechanisms \cite{Ijjas:2015zma, Moffat:2014poa,Brandenberger:2011et, Poplawski:2010kb, Khoury:2001wf}.

One such proposed mechanism is the time-delayed cosmology of Choudhury et al. \cite{Choudhury:2011xf}. In this proposal, the evolution of the energy density of the Universe as expressed in the Friedmann equation is delayed by a constant $\tau$ relative to expansion. While ad hoc and somewhat unnatural, the scheme does generate inflation without the aforementioned problems of inflation models. An exploration of its consequences is premised on the possibility of some nonlocal theories effectively generating time-delayed responses in gravitational dynamics \cite{Ng:2008pi, Markopoulou:2007ha, Eliezer:1989cr}, and on the richer dynamics afforded by time-delayed systems  \cite{Smith:2011,Erneux:2009}. More broadly, it answers a general invitation to explore the potential role of delay differential equations in fundamental physics \cite{Atiyah:2010qs}.

However, time-delayed cosmology has received scant attention from the community, perhaps largely owing to its detachment from fundamental theory. Not surprisingly, there are hardly any observational constraints on its key parameter, the time delay $\tau$, which is generally just presumed to be of the order of the Planck time. One notable attempt was made in Ref. \cite{Yang:2014aua}, where the matter power spectrum was calculated in an ad hoc general relativistic delay scheme. It also obtained an estimate for the delay parameter, though this was not based on the observational power spectrum data.

In this work, we adopt the stance that the time-delayed cosmology proposal can at least be empirically interesting and make initial steps towards filling the aforementioned constraint gap. In order to do this, we apply time-delayed cosmology to the late Universe where an optically invisible fluid, often dubbed dark energy, supersedes matter and radiation to source the observed late-time cosmic acceleration \cite{SupernovaCosmologyProject:1998vns, SupernovaSearchTeam:1998fmf, WMAP:2003elm, Pan-STARRS1:2017jku, Planck:2018vyg}. The substantial evidence for dark energy and the theoretical parallels between primordial inflation and dark energy make the application of time-delayed cosmology to the dark Universe today worth undertaking.

In particular, we consider the effects of the delayed Friedmann equation [Eq. \eqref{eqn:delay_a}] at late times and determine the background evolution as well as the growth of Newtonian perturbations about this delayed background expansion. We calculate the Hubble expansion rate $H(z)$ as well as two growth observables, the growth rate $f(z)$ and $f\sigma_8(z)$. We show clear dependence of the predictions on the time delay parameter $\tau$ and estimate this parameter directly from observational data. This hints at the potential relevance of time-delayed cosmology in a cosmic era that has not been demonstrated until now.

In the next section, we briefly introduce important details of time-delayed cosmology. In Section \ref{background}, we discuss the background evolution. In Section \ref{perturbations}, we set up the equations for the growth of matter perturbations and discuss the predictions of time-delayed cosmology. In Section \ref{estimate}, we perform a Markov chain Monte Carlo sampling and estimate the time delay parameter $\tau$ directly from the Hubble expansion rate $H(z)$, the growth rate $f(z)$, and $f\sigma_8(z)$ data. Finally, we conclude our work in Section \ref{conclusion}. The code for reproducing the figures and calculations in this paper can be freely downloaded at \cite{testing_time_delay}.

\textit{Conventions.} We work with the geometrized units $c = 8 \pi G = 1$ and the mostly plus metric signature $(-,+,+,+)$. A dot over a variable denotes differentiation with respect to the cosmic time $t$.

\section{Time-delayed cosmology}\label{delayed_cosmology}

We provide a short introduction to the foundations of time-delayed cosmology (Section \ref{subsec:foundations}) and discuss the method of steps for solving a delay differential equation (Section \ref{subsec:method_of_steps}). We then describe the set-up of time-delayed cosmology at late times (Section \ref{subsec:application}). 

\subsection{Foundations}
\label{subsec:foundations}

With the observational support for large-scale statistical homogeneity \cite{Goncalves:2020erb, Ntelis:2017nrj, Li:2015yha} and isotropy \cite{Sarkar:2018smv, Saadeh:2016sak, Hazra:2015wla}, the standard description of cosmic evolution is given by the following Friedmann equation:
\begin{equation}
    \left(\dfrac{\dot{a}(t)}{a(t)}\right)^2 = \dfrac{1}{3}\rho(t),
\end{equation}
where $a(t)$ is the scale factor which measures the expansion of the cosmos and $\rho(t)$ is the energy density of the perfect fluid permeating the Universe. Assuming an equation of state of the form $p(t) = \omega \rho(t)$ (where $p$ is the fluid pressure and $\omega$ is a constant called the equation of state parameter) and solving the continuity equation,
\begin{equation}\label{friedmann_equation}
    \dot{\rho}(t) = -3(\rho(t) + p(t))\dfrac{\dot{a}(t)}{a(t)},
\end{equation}
the Friedmann equation can be written as
\begin{equation}
    \left(\dfrac{\dot{a}(t)}{a(t)}\right)^2 = \dfrac{\rho_{i,x}}{3}a(t)^{-3(1+\omega)},
\end{equation}
where $\rho_{i,x}$ is the initial energy density of the fluid denoted by $x$. A late Universe described by the Friedmann equation and filled with a mixture of cold (i.e. pressureless) dark matter ($\omega = 0$) and a cosmological constant $\Lambda$ ($\omega = -1$) is referred to as the standard $\Lambda$CDM model. 

On the other hand, time-delayed cosmology is based on a delayed Friedmann equation
\begin{align}
    \left(\dfrac{\dot{a}(t)}{a(t)}\right)^2 &= \dfrac{\rho_{i,x}}{3}a(t-\tau)^{-3(1+\omega)}\label{eqn:delay_a},
\end{align}
where $\tau$ is some constant that, in the original application in the inflationary era, has units of Planck time $t_p \sim \mathcal{O}(10^{-44})$ s. Delaying the energy density term in this way is completely ad hoc and not unique. 

A fundamental action is yet to be found to prescribe this time-delayed set-up. However, there are motivations for the delay. A nonlocal theory of (quantum) gravity may induce a delayed response on the Universe since nonlocality may imply time-smeared interactions \cite{Choudhury:2011xf}. For example, the Deser-Woodard models \cite{Deser:2019lmm,Deser:2007jk,Chen:2019wlu}, which are inspired by quantum loop corrections, involve cosmological equations with retarded boundary conditions. A more explicit example is Ref. \cite{Beaujean:2009rb} in which nonlocality has resulted to equations of motion that are systems of delay differential equations. In the application to the late-time era, we take time-delay to be a phenomenological step to alternatively source cosmic acceleration. We shall not expound on the origins of the delay, rather we shall regard this delay as a phenomenological parameter. 

\subsection{The method of steps}
\label{subsec:method_of_steps}

\begin{figure}[h!]
    \centering
    \includegraphics[width=1\linewidth]{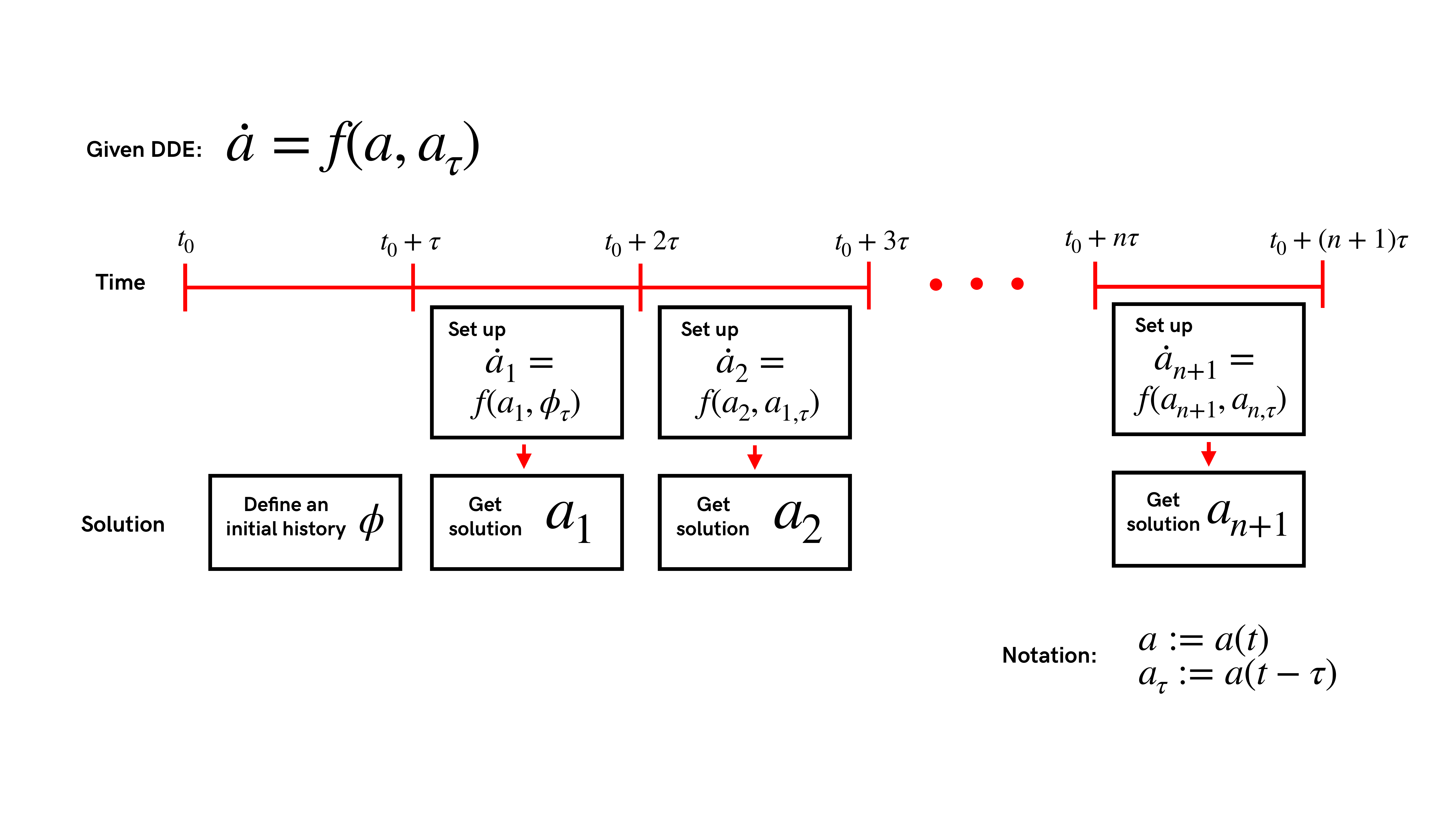}
    \caption{The method of steps. Starting with a given delay differential equation, we define an initial history $\phi(t)$ on an interval at least the size of one delay unit, e.g. $[t_0, t_0 + \tau)$. On the succeeding interval $[t_0 + \tau, t_0 + 2\tau)$, we replace the delayed term with $\phi(t-\tau)$ and solve the ensuing \textit{ordinary} differential equation, using $\phi(t_0 + \tau)$ as an initial value. We label the solution of this ordinary differential equation as $a_1(t)$. On the following interval $[t_0 + 2\tau, t_0 + 3\tau)$, again with a size of one delay unit,  we replace the delayed term with $a_1(t-\tau)$ and again solve the ensuing ordinary differential equation for $a_2(t)$, with $a_1(t_0 + 2\tau)$ as an initial value. We repeat this process for as many intervals as we like, using the previous solution $a_n(t)$ to replace the delayed term and solve the resulting ordinary differential equation for $a_{n+1}(t)$. The piecewise function defined by $a_n(t)$'s comprise the solution to the delay differential equation.}
    \label{fig:method_of_steps}
\end{figure}

The delayed Friedmann equation can be solved with the method of steps \cite{Smith:2011, Erneux:2009}. The essential idea of the method of steps is to replace the delayed term with a known solution $a(t)$ so that the delay differential equation becomes ordinary within an interval that is then solvable with standard methods. Effectively, the solution to the delayed equation (as with any constant-delay differential equation) is a piecewise function with each composite solution defined on an interval of the size of the delay $\tau$. The first composite solution, which is to be defined, is called an initial history. This is the equivalent of the initial value in ordinary differential equations. 
Figure \ref{fig:method_of_steps} illustrates the method of steps. 

\begin{figure}[h!]
    \centering
    \includegraphics[width=0.6\linewidth]{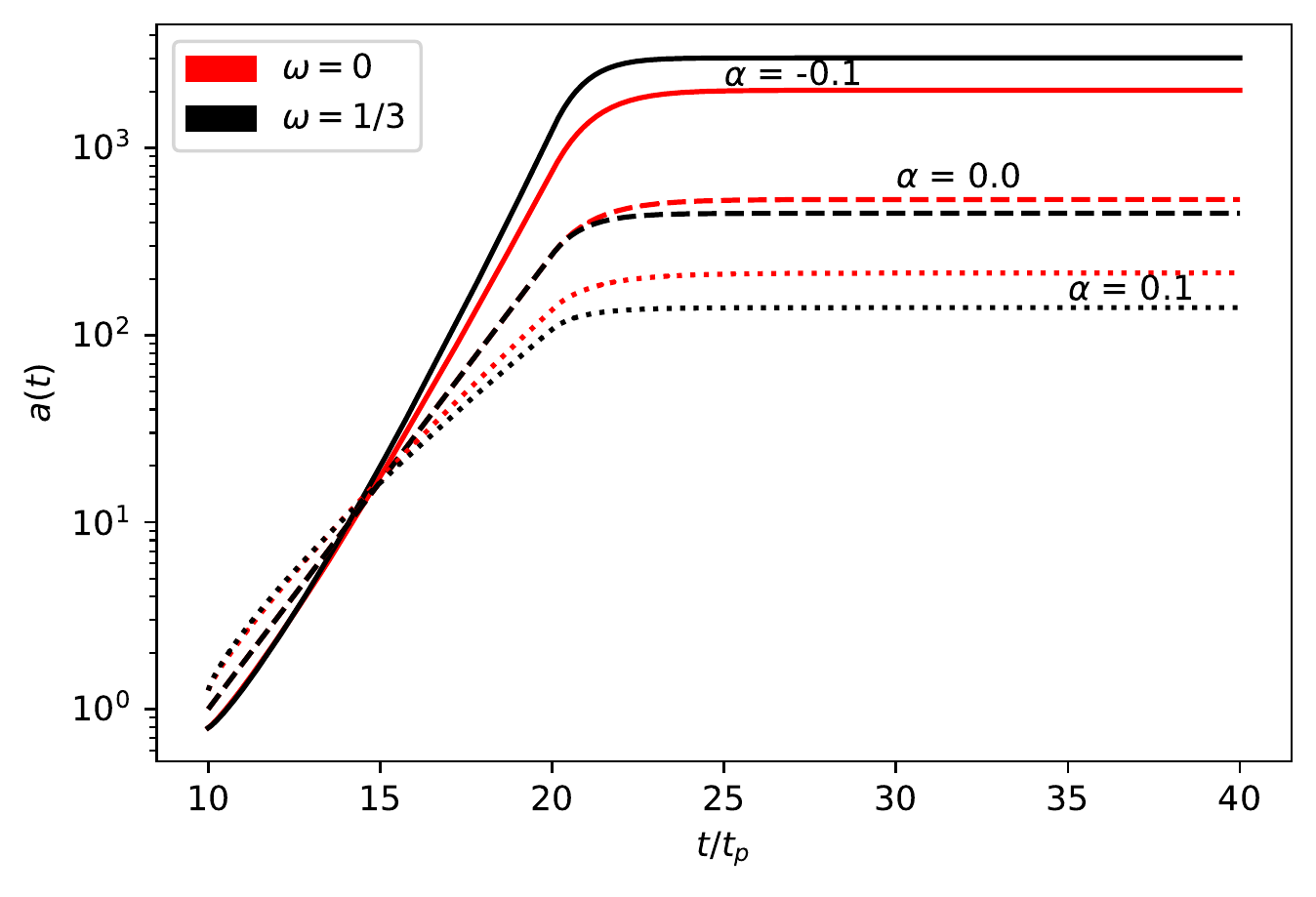}
    \caption{ Solution to the delayed Friedmann equation due to a delay $\tau = 10t_p$, where $t_p$ is Planck time. The inflationary period lasts for one delay unit (in this figure, from $t = 10 t_p$ to $t =  20 t_p$) before the scale factor transitions to a decelerated evolution. }
    \label{fig:inflationary_solution}
\end{figure}

For a power-law initial history $\phi(t) = t^\alpha$ defined on $t \in [0, \tau)$, which may be due to a quantum gravity effect or a pre-Big Bang scenario, the delayed Friedmann equation admits inflation in the following interval:
\begin{align}
    a(t) &= \phi(\tau)\exp\left(\sqrt{\dfrac{\rho_{i,x}}{3}}\dfrac{(t-\tau)^\gamma}{\gamma}\right), \ t \in [\tau, 2\tau),\\
    \gamma &= 1-\dfrac{3}{2}(1+\omega)\alpha.
\end{align}
For succeeding times, the  delayed equation has to be solved numerically. In this work, we use the \verb|ddeint| Python package \cite{ddeint} for the numerical solutions. Clearly, in time-delayed cosmology, inflation can be naturally generated for a period of one delay unit. This period also seamlessly ends, thereby avoiding the ``graceful exit'' problem (see Figure \ref{fig:inflationary_solution}).

\subsection{Application to late times}\label{subsec:application}

\begin{figure}[h!]
    \centering
    \includegraphics[width=0.6\linewidth]{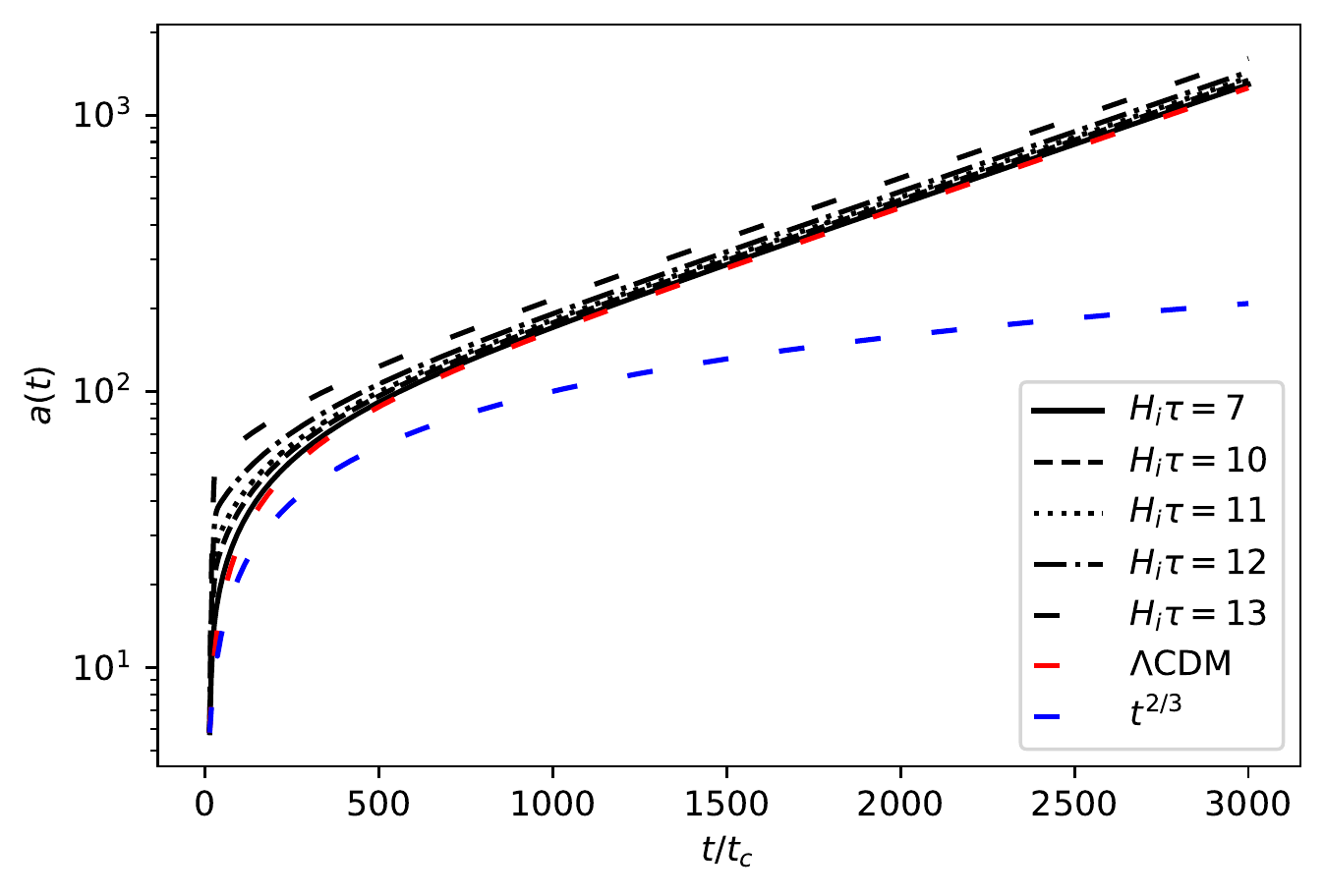}
    \caption{Solution to the time-delayed Friedmann equation in the presence of a cosmological constant and time delay on the order of $t_c$, where $t_c = H_i^{-1} = 0.0175 \pm 0.0001$ Gyr. The integration starts off with an initial history of $\phi(t) \sim t^{2/3}$ in the matter-dominated era and evolves into the future.}
    \label{fig:cosmic_acceleration}
\end{figure}

At late times, the phenomenon of interest is cosmic acceleration due to dark energy. Because of the parallels between primordial inflation and late-time cosmic acceleration, the application of the delayed Friedmann equation at late times is worth considering. Furthermore, it will be easier to place constraints on time-delayed cosmology if we can show its impact on the expansion era. 

In this application, we phenomenologically regard dark energy as a mixture of a cosmological constant and a time delay. The delayed Friedmann equation is therefore of the form
\begin{equation}\label{eqn:friedmann_hi}
    H(t)^2 = H_i^2\left(\dfrac{\Omega_{i,m}}{a^3(t-\tau)}+\Omega_{i,\Lambda}\right),
\end{equation}
where $H(t):= \dot{a}(t)/a(t)$ is the Hubble parameter, $H_i$ is the initial Hubble parameter value, $\Omega_{i,m} := \rho_{i,m}/(3H_i^2)$ is the matter density parameter, and $\Omega_{i,\Lambda} := \rho_{i,\Lambda}/(3H_i^2)$ is the cosmological constant density parameter. Figure \ref{fig:cosmic_acceleration} shows that the delayed Friedmann equation can also accommodate a late-time cosmic acceleration. 

Note that in the original implementation of the delay modification in Ref. \cite{Choudhury:2011xf}, as well as in Ref. \cite{Yang:2014aua}, the delay $\tau$ is assumed to have units of Planck time $t_p$, being the relevant time scale for inflation. Because the delay is very small, it would have no impact on late-time observables, which is why the computed power spectrum in Ref. \cite{Yang:2014aua} expectedly appears to be in excellent agreement with observations. In this application, we allow the delay to be large since the relevant time scale $t_c$ is also large; we will integrate in the matter-dominated era up to the present dark energy-dominated era. We will solve the  dimensionless version of Equation $\ref{eqn:friedmann_hi}$ and the relevant time scale would be $t_c = H_i^{-1}$, with $H_i$ being the Hubble parameter in the matter-dominated era. We find the value of $H_i$ using the following relation for a \textit{constant} dark energy density
\begin{equation}\label{eqn:de_1}
    3H_i^2\Omega_{i,\Lambda} = 3H_0^2\Omega_{0,\Lambda},
\end{equation}
where $H_0$ and $\Omega_{0, \Lambda}$ are the Hubble constant and the present value of the cosmological constant density parameter, respectively. The latest Planck 2018 estimates are $H_0 = 67.4 \pm 0.5$ km s$^{-1}$ Mpc$^{-1}$ and $\Omega_{0,\Lambda} = 0.6889 \pm 0.0056$ \cite{Planck:2018vyg}. Solving for $H_i$ in Equation \ref{eqn:de_1}, we get
\begin{equation}
    H_i = H_0\sqrt{\dfrac{\Omega_{i,\Lambda}}{\Omega_{0, \Lambda}}}.
\end{equation}
 Starting our integrations in the time when $\Omega_{i,\Lambda} = 10^{-6}$ (and $\Omega_{i,m} =  1 - \Omega_{i, \Lambda}$), this results to a time scale of $t_c = 0.0175 \pm 0.0001 $ Gyr, where here and throughout the paper we have used the \verb|SOAD| package \cite{Erdim:2019nzq} for (asymmetric) error propagation. The units of the time delay will be in terms of this time scale $t_c$. 

\section{Hubble expansion}\label{background}

\begin{figure}
    \centering
    \includegraphics[width=0.6\linewidth]{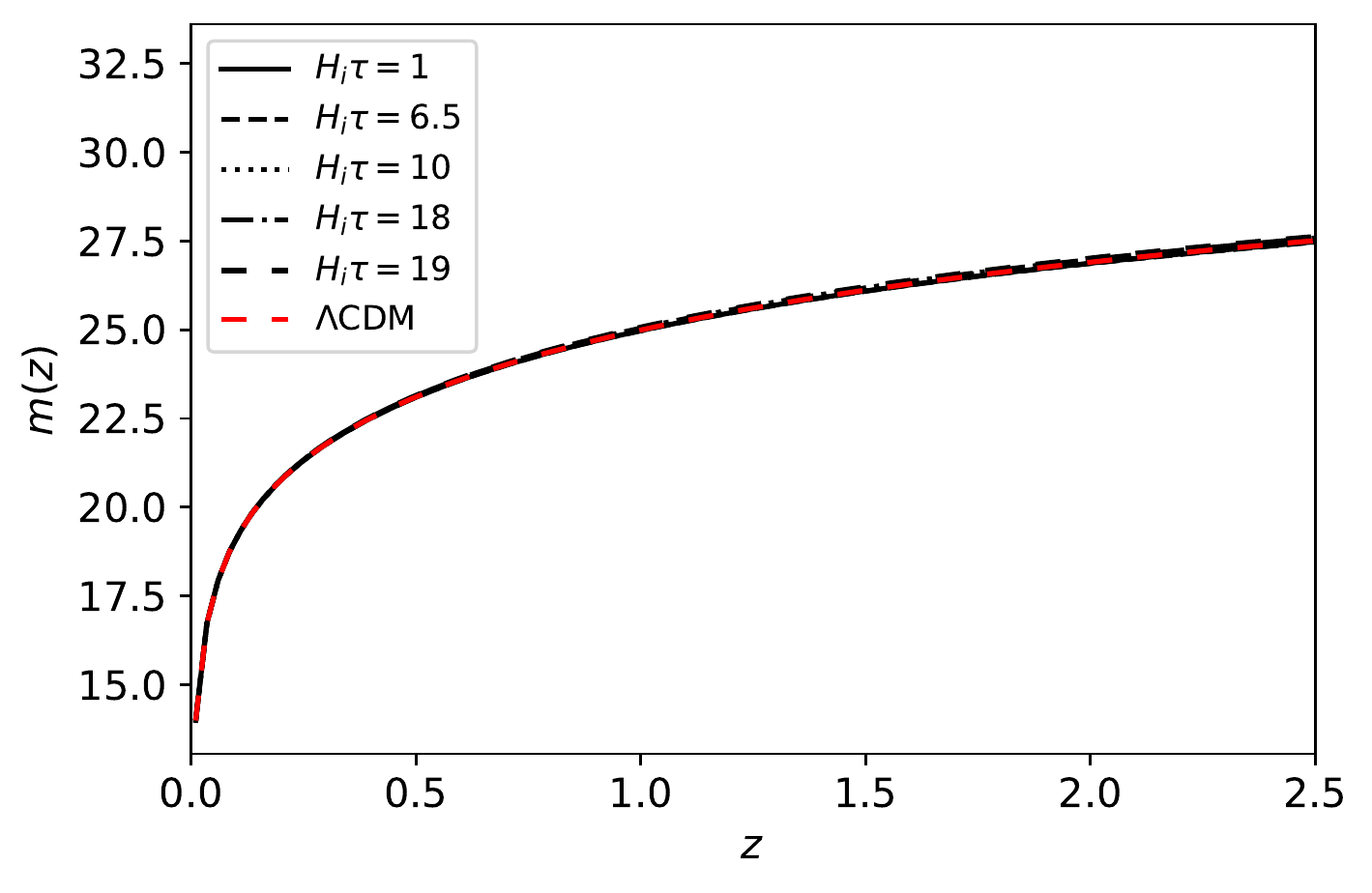}
    \caption{The evolution of the apparent magnitude $m(z)$ of Type Ia supernovae. Here, we have assumed that the absolute magnitude takes the value $M = -19.3$ and $H_0$ is given by the Planck 2018 estimate. Time-delayed cosmology is virtually indistinguishable from $\Lambda$CDM.}
    \label{fig:appmag_vartau_full}
\end{figure}

To obtain the background evolution from Equation \ref{eqn:friedmann_hi}, we must specify an initial history. Throughout this paper, we assume a power-law initial history of the form $\phi(t) \sim t^\alpha$ for the delayed Friedmann equation. We have checked numerically that the observables we are interested in in this paper do not strongly depend on the parameter $\alpha$ (see Figure \ref{fig:all_varalpha_close}) in redshifts that are currently accessible to us and especially for reasonable values of $\alpha$ (that is, for $\alpha \approx 2/3$ which refers to the canonical matter-era solution). Furthermore, although $\alpha$ gains a stronger effect at very large redshifts in terms of affecting the magnitude of the observables and for very large time delays ($H_i\tau > 10$), the general shape of the observables are determined by the time delay parameter and not by $\alpha$. For these reasons and combined with the fact that an $\alpha = 2/3$ constitutes a more natural initial history (taking after the canonical $t^{2/3}$ matter-era solution), we choose to fix the value of $\alpha$ to 2/3 in the following calculations instead of taking it as a free parameter.

\begin{figure}
    \centering
    \includegraphics[width=0.6\linewidth]{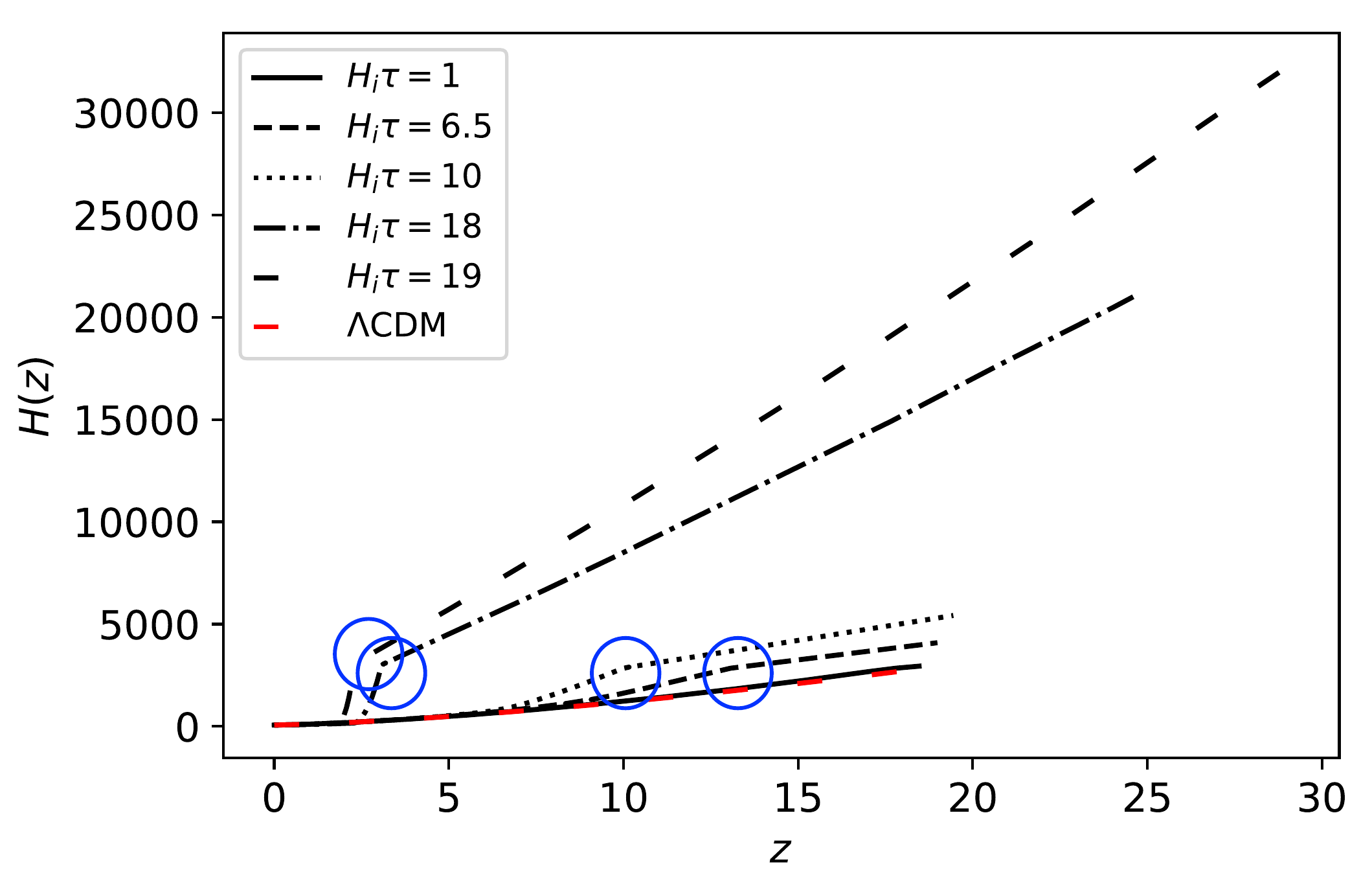}
    \caption{The evolution of the Hubble expansion rate $H(z)$. The predictions have been normalized to have the value of the Planck 2018 estimate of the Hubble constant $H_0$ at $z = 0$. A striking feature of time-delayed cosmology is a kink or a point at which its prediction changes sharply. These kinks are encircled in blue above. }
    \label{fig:hubble_vartau_full}
\end{figure}

Due to Figure \ref{fig:cosmic_acceleration}, we can already expect that the background evolution of time-delayed cosmology closely follows that of $\Lambda$CDM. Indeed, if we look at the predictions for the apparent magnitude $m(z)$ of supernovae in Figure \ref{fig:appmag_vartau_full}, we can see that time-delayed cosmological predictions are virtually indistinguishable from the $\Lambda$CDM prediction. This is the case even when considering delays on the order of $H_i\tau \sim 10$ and when considering larger redshifts.  The difference between $\Lambda$CDM and time-delayed cosmology is revealed when we look at the Hubble expansion rate $H(z)$. Figure \ref{fig:hubble_vartau_full} shows the evolution of the Hubble expansion rate $H(z)$ for a fixed $H_0$. The dashed red curve shows the prediction of the standard $\Lambda$CDM model and the black curves are the predictions of time-delayed cosmology. Predictions due to delays that are larger than $H_i\tau = 1$ already notably deviate from the $\Lambda$CDM prediction at redshifts $z > 5$. Notice that the predictions appear to start at different redshifts. This is the case for all the redshift plots in this paper. This happens because different delays affect the scale factor evolution differently, which is then used to obtain the redshift. However, we have made sure that all quantities start out with the same initial condition at the same starting integration time.

A striking observation in Figure \ref{fig:hubble_vartau_full} are kinks (encircled in blue) or points at which the predictions of time-delayed cosmology change sharply. In fact, the derivative of $H(z)$ at any of the kinks is undefined. This is an expected feature and an artifact of delay differential equation models. It is well known that discontinuities propagate in the derivatives of the solution to delay differential equations \cite{Smith:2011, Erneux:2009}. At the start of integration, the first derivative is discontinuous. One delay unit afterwards, the discontinuity propagates in the second derivative. Since our definition of the Hubble expansion rate involves $\dot{a}(t)$, we expect to see the discontinuity in the second derivative $\Ddot{a}(t)$ at a certain point (the kinks). 

\begin{figure}
    \centering
    \subfloat{\makebox[0.5\textwidth]{{\includegraphics[width=0.6\linewidth]{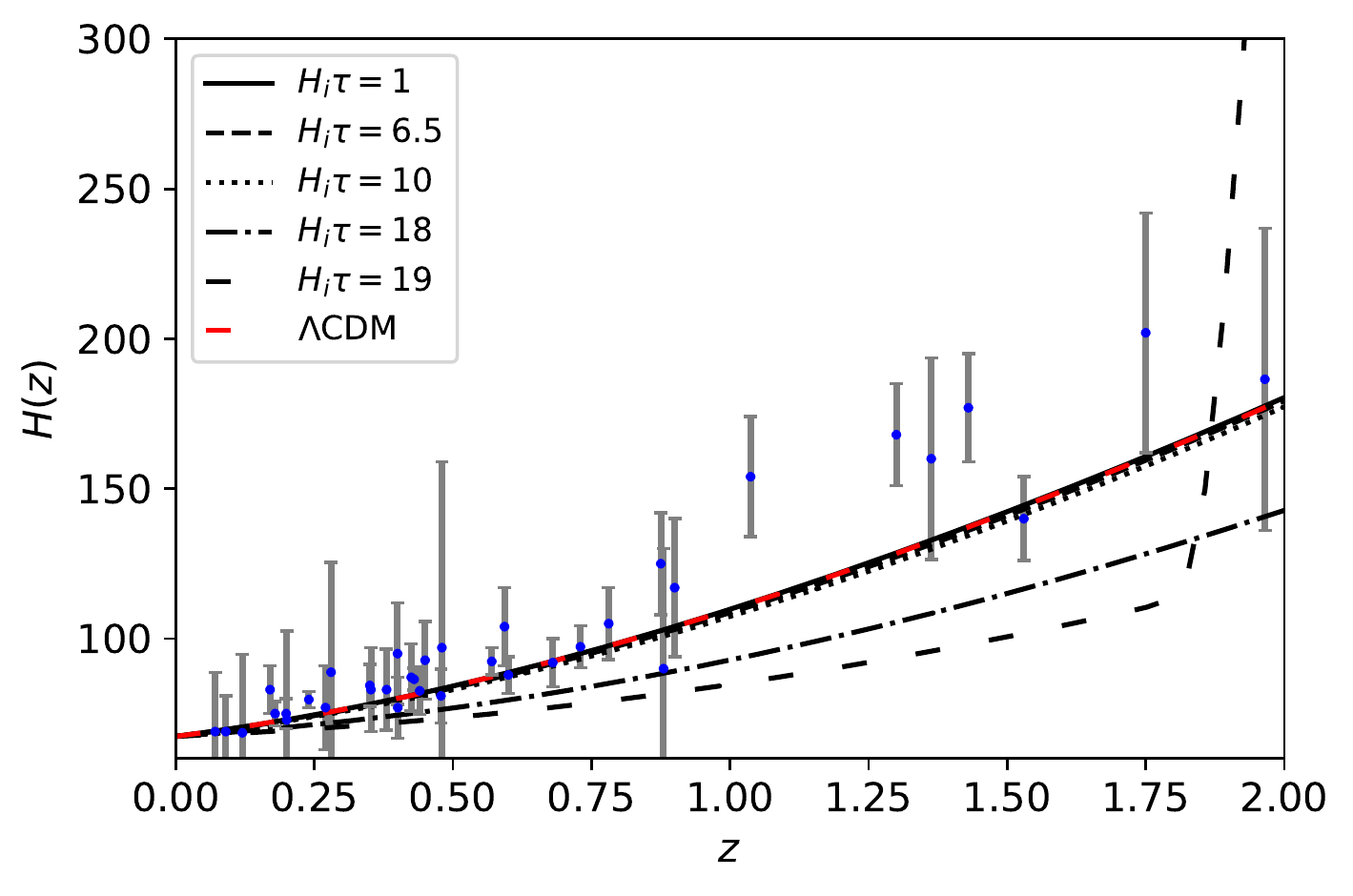}}}}
    \quad
    \subfloat{\makebox[0.5\textwidth]{{\includegraphics[width=0.6\linewidth]{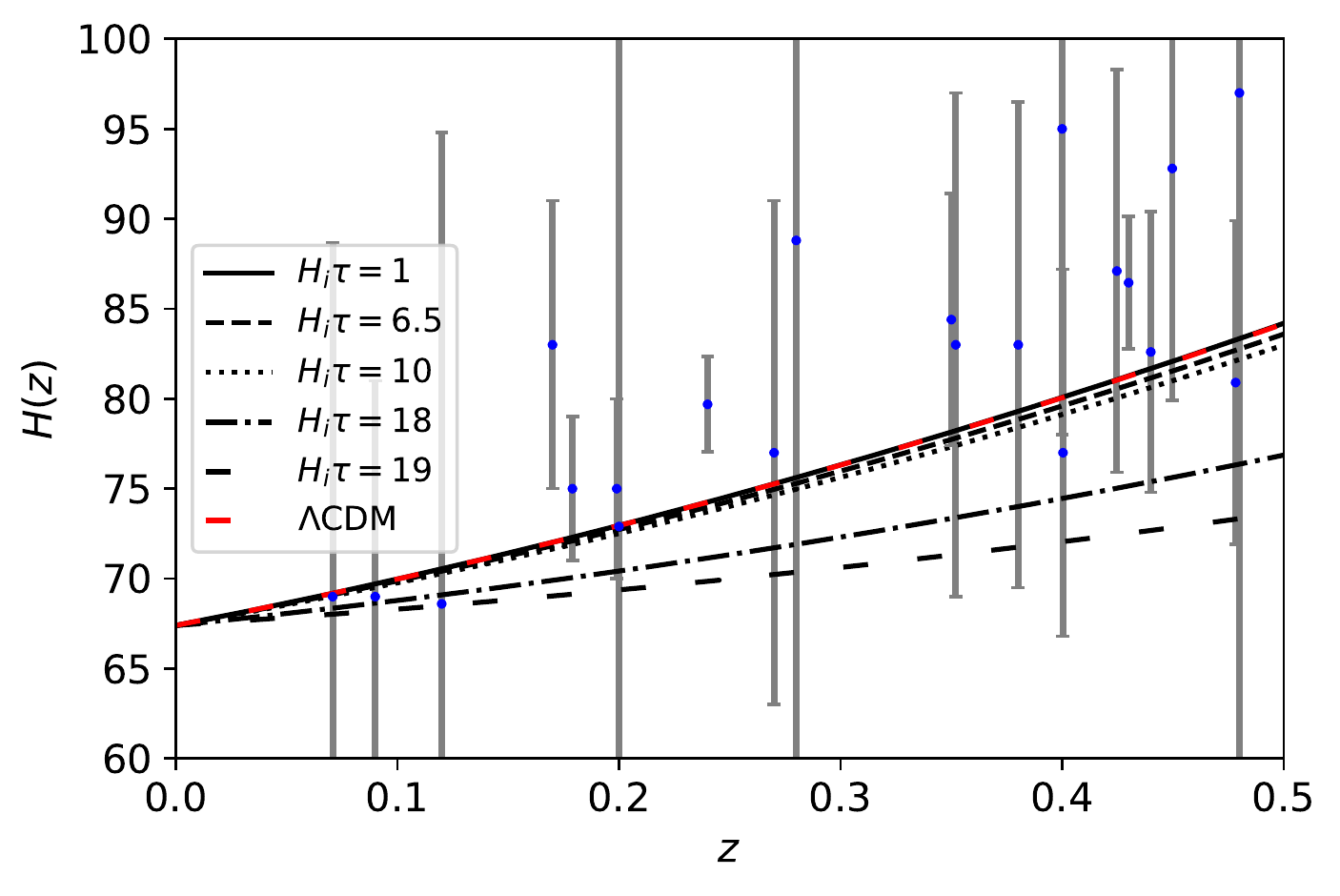}}}}
    \quad
    \caption{The evolution of the Hubble expansion rate $H(z)$ at small and intermediate redshifts. Time-delayed cosmology closely follows $\Lambda$CDM even for delays on the order of $H_i\tau \sim 10$. The Hubble data are taken from the compilation in Ref. \cite{Gogoi:2021mhi}.}\label{fig:hubble_vartau_close}
\end{figure}

These kinks in the Hubble expansion rate already provide an upper bound on the time delay without further statistical analysis. In Figure \ref{fig:hubble_vartau_close}, we can see that a time delay with magnitude $H_i\tau \approx 19$ can already be ruled out due to the presence of a kink that the data clearly does not accommodate. As the value of the time delay is increased, the kinks in the Hubble expansion rate are revealed at smaller and smaller redshifts. This is also true if we consider other observables. Therefore, all time delays $H_i\tau > 19$ are also ruled out.

Of course, we do not expect real, physical quantities of the Universe to exhibit these discontinuities. But if our Universe is correctly modeled by a delayed Friedmann equation, the abrupt transitions in $H(z)$ serve as a generic smoking gun that make them empirically interesting. There may be fundamental reasons behind these discontinuities. For example, the improved Deser-Woodard model \cite{Deser:2019lmm} was shown to have a discontinuous evolution of matter perturbation \cite{Ding:2019rlp}. A possible cause of the discontinuity is a strong nonlocal effect.

For time delays with magnitude $H_i\tau < 19$, the smoking-gun imprints of time-delayed cosmology on the background evolution are only revealed at large redshifts for which data is still unavailable. When we look at redshifts $z < 2$ (Figure \ref{fig:hubble_vartau_close}), we can see that time-delayed cosmology closely follows $\Lambda$CDM even for delays on the order of $H_i\tau \sim 10$. This shows that the key time delay parameter does not have to be of the order of Planck time as originally envisaged in order to fit observational data. Even large cosmic delays appear to be viable. An unfortunate consequence of this is that the Hubble expansion data is unable to distinguish time-delayed cosmology from $\Lambda$CDM. To observe the difference, we look to Newtonian perturbations.

\section{Newtonian perturbations}\label{perturbations}

In this work, we choose the growth of Newtonian matter perturbations as an additional probe of time-delayed cosmology. In particular, we are interested in the growth rate $f(z)$ and another observable $f\sigma_8(z)$, where $z$ is the redshift, that are both dependent on the amplitude of perturbations. The former quantity is the speed of growth of perturbations in the Universe with respect to the cosmic expansion, and the latter is essentially the growth rate scaled by the evolving root-mean-square of matter perturbations. These observational probes of large-scale structures have been used to distinguish between modified gravity theories and the standard $\Lambda$CDM model \cite{Huterer:2013xky, Amendola:2003wa,LeviSaid:2021yat, Perenon:2019dpc, Kazantzidis:2018rnb, Linder:2018jil, Hirano:2010yf, Mirzatuny:2013nqa, Basilakos:2012uu}. As we shall see, these will also be useful for obtaining constraints on time-delayed cosmology.

\subsection{Set-up}

The growth of Newtonian matter perturbations is given by \cite{Huterer:2013xky, Amendola:2003wa}
\begin{equation}
\label{eq:newtonian_fluctuation_eq}
    \Ddot{\delta}(t) + 2H(t)\dot{\delta}(t) - \rho_m(t)\delta(t) = 0,
\end{equation}
where $\delta(t) := \delta\rho_m(t)/\rho_m(t)$ is the density contrast quantifying the inhomogeneity of the universe and $\rho_m(t)$ is the background energy density of (dark) matter. This fluctuation equation is valid for sub-horizon perturbations, i.e., $\lambda/ a(t) \ll H(t)^{-1}$, where $\lambda$ is the co-moving mode wavelength of the density perturbation. It is convenient to rewrite and solve this equation in terms of the scale factor $a$ or the redshift $z$ (using the relation $z = (1/a) - 1$). Once $\delta(a)$ is obtained, the two observables of interest can be easily calculated using the following definitions:
\begin{align}
    f(a) &:= \dfrac{d\ln{\delta(a)}}{d\ln{a}},\\
    f\sigma_8(a) &:= \dfrac{\sigma_8}{\delta(a=1)}\delta(a)f(a),
\end{align}
where $\sigma_8$ is the present root-mean-square variance in the number of galaxies in spheres of radius 8$h^{-1}$ Megaparsec (with $h = H_0/(100$ km s$^{-1}$ Mpc$^{-1}$) being the dimensionless value of the Hubble parameter today). Note that these two observables are fundamentally independent quantities; whereas $f(a)$ carries information on $d\delta/da$, $f\sigma_8(a)$ carries information on $\delta(a)$.

In what follows, we solve the fluctuation equation
\begin{equation} \label{eqn:fluc}
    \Ddot{\delta}(t) + 2H(t)\dot{\delta}(t) - \dfrac{3}{2}H^2(t)\delta(t) = 0,
\end{equation}
where we have replaced the background matter energy density with its Hubble function equivalent using the (delayed) Friedmann equation. The observables of interest in time are then given by
\begin{align}
    f(t) &:= \dfrac{d\ln{\delta(t)}}{d\ln{a(t)}} = \dfrac{1}{H(t)}\dfrac{\dot{\delta}(t)}{\delta(t)},\\
    f\sigma_8(t) &:= \dfrac{\sigma_8}{\delta(t=t_0)}\dfrac{1}{H(t)}\dot{\delta}(t),
\end{align}
where $t_0$ denotes the present day. In addition to assuming an initial history of the form $\phi(t) \sim t^{2/3}$ for reasons we mentioned before, we also set the canonical $a(t) \sim \delta(t) \sim t^{2/3}$ solution as an initial condition for the perturbation equation. This means that we integrate deep in the matter-dominated era up to the present dark energy-dominated era. In comparing the results with the $\Lambda$CDM model, we use the latest value of $\sigma_8$ given by Planck: $\sigma_8 = 0.811\pm 0.006$ \cite{Planck:2018vyg}.

We note that we are using the standard (i.e. non-delayed) perturbation equation here instead of a new delayed perturbation equation. Absent a fundamental action for time-delayed cosmology, this is unfortunately the best that one can do, short of proposing further ad hoc prescriptions about how the delay directly affects perturbations. Here, we  adopt the conservative position that time-delayed cosmology manifests itself only through the cosmic expansion, i.e. the Hubble function. We find that this conservative modification is enough to see interesting consequences of time-delayed cosmology without having to develop an action-based delayed perturbation theory. 

\subsection{Growth rate $f(z)$}

\begin{figure}
    \centering
    \includegraphics[width=0.6\linewidth]{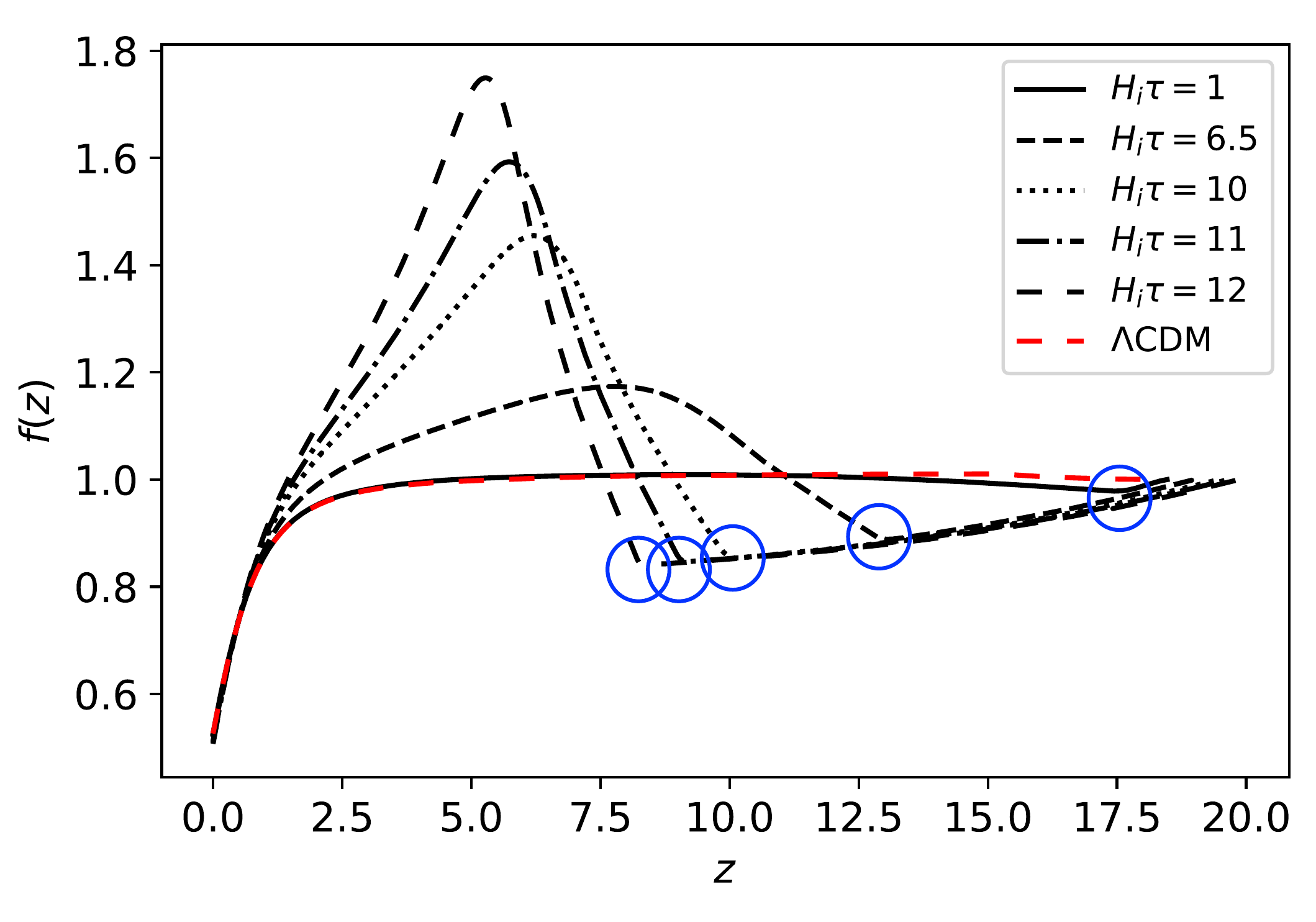}
    \caption{The evolution of the growth rate $f(z)$. Time-delayed cosmology with intermediate (i.e. $H_i\tau \sim 10$) delay parameter values predict markedly different growth rate evolutions. In particular, the predictions for time-delayed cosmology decreases initially before increasing and eventually peaking. There are also kinks (encircled in blue) in the predictions.}
    \label{fig:growth_rate_vartau_full}
\end{figure}

\begin{figure}
    \centering
    \subfloat{\makebox[0.5\textwidth]{{\includegraphics[width=0.6\linewidth]{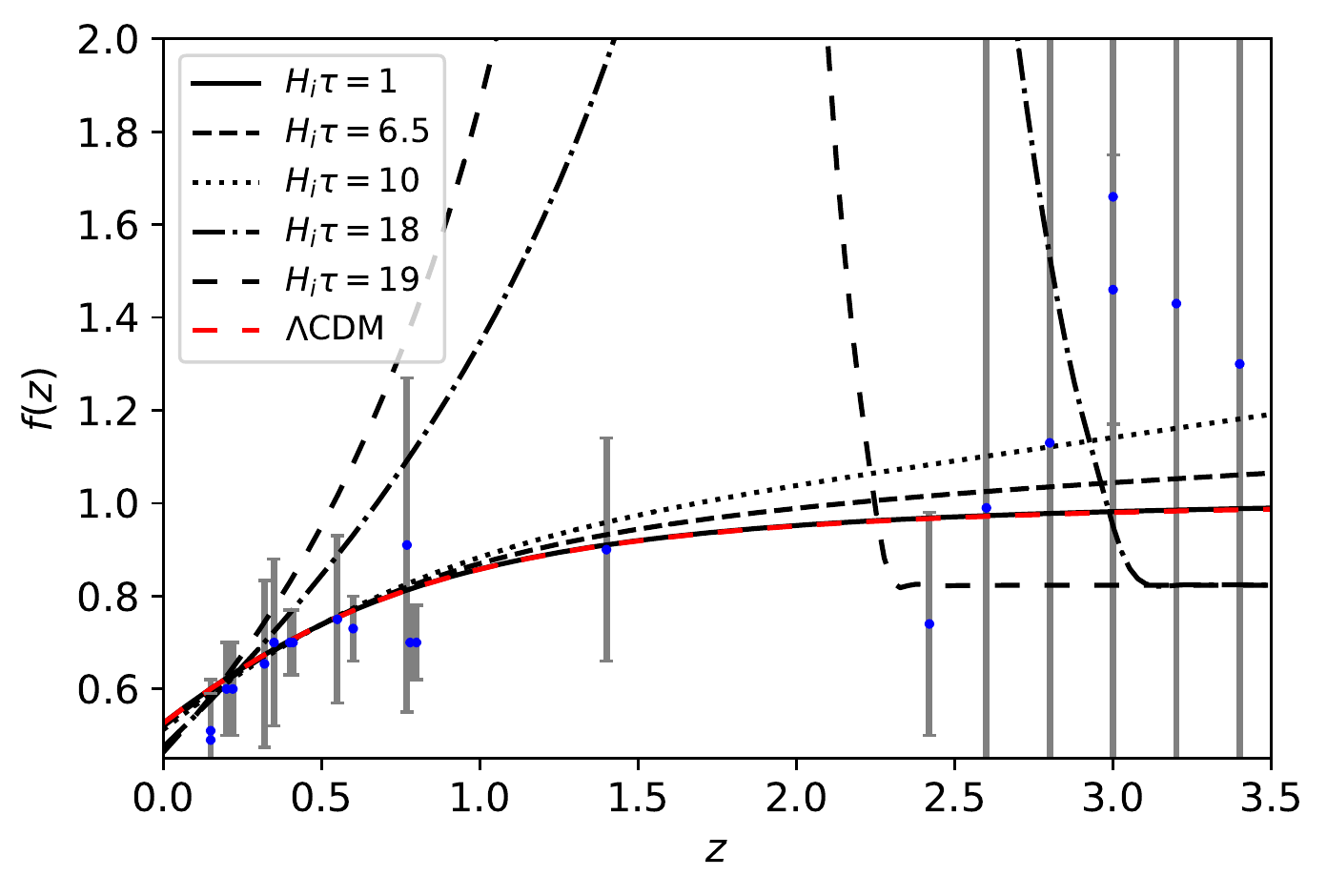}}}}
    \quad
    \subfloat{\makebox[0.5\textwidth]{{\includegraphics[width=0.6\linewidth]{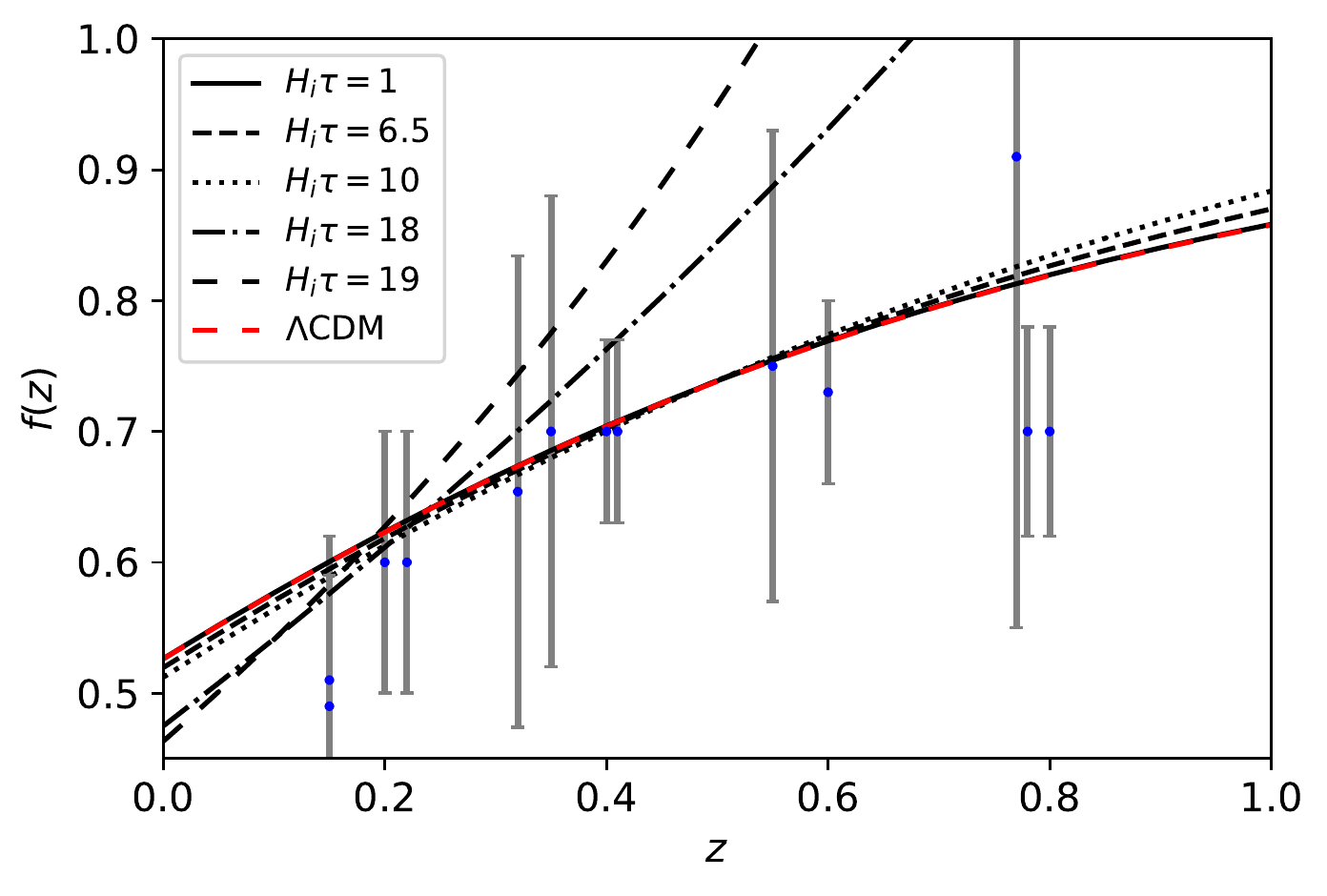}}}}
    \quad
    \caption{The evolution of the growth rate $f(z)$ at small and intermediate redshifts. The growth rate data are taken from the compilation in Refs. \cite{Mirzatuny:2013nqa} and \cite{Peel:2012vg}.}\label{fig:growth_rate_vartau_close}
\end{figure}

Figure \ref{fig:growth_rate_vartau_full} shows the plot of the growth rate $f(z)$. The dashed red curve shows the prediction of the standard $\Lambda$CDM model, whereas the black curves are the predictions of time-delayed cosmology at different values of the time delay parameter $\tau$. Immediately, we can see that time-delayed cosmology makes very different predictions for $f(z)$ even for ``intermediate'' (i.e. $H_i\tau \sim 10$) values of the time delay parameter. The growth rate of a delayed Universe dips in the matter-dominated era (i.e. $z > 1$) and then peaks later on before dark energy finally suppresses it for good (see Figure \ref{fig:growth_rate_vartau_close}). On the other hand, if the delays are ``small'' (i.e. $H_i\tau \lesssim 1$), then these dips and peaks are weak or not visible at all and the growth rate is virtually indistinguishable from the prediction of $\Lambda$CDM.

\begin{figure}
    \centering
    \subfloat[Ratio of the density contrast to the scale factor\label{fig:delta_over_a}]{\makebox[0.5\textwidth]{{\includegraphics[width=0.6\linewidth]{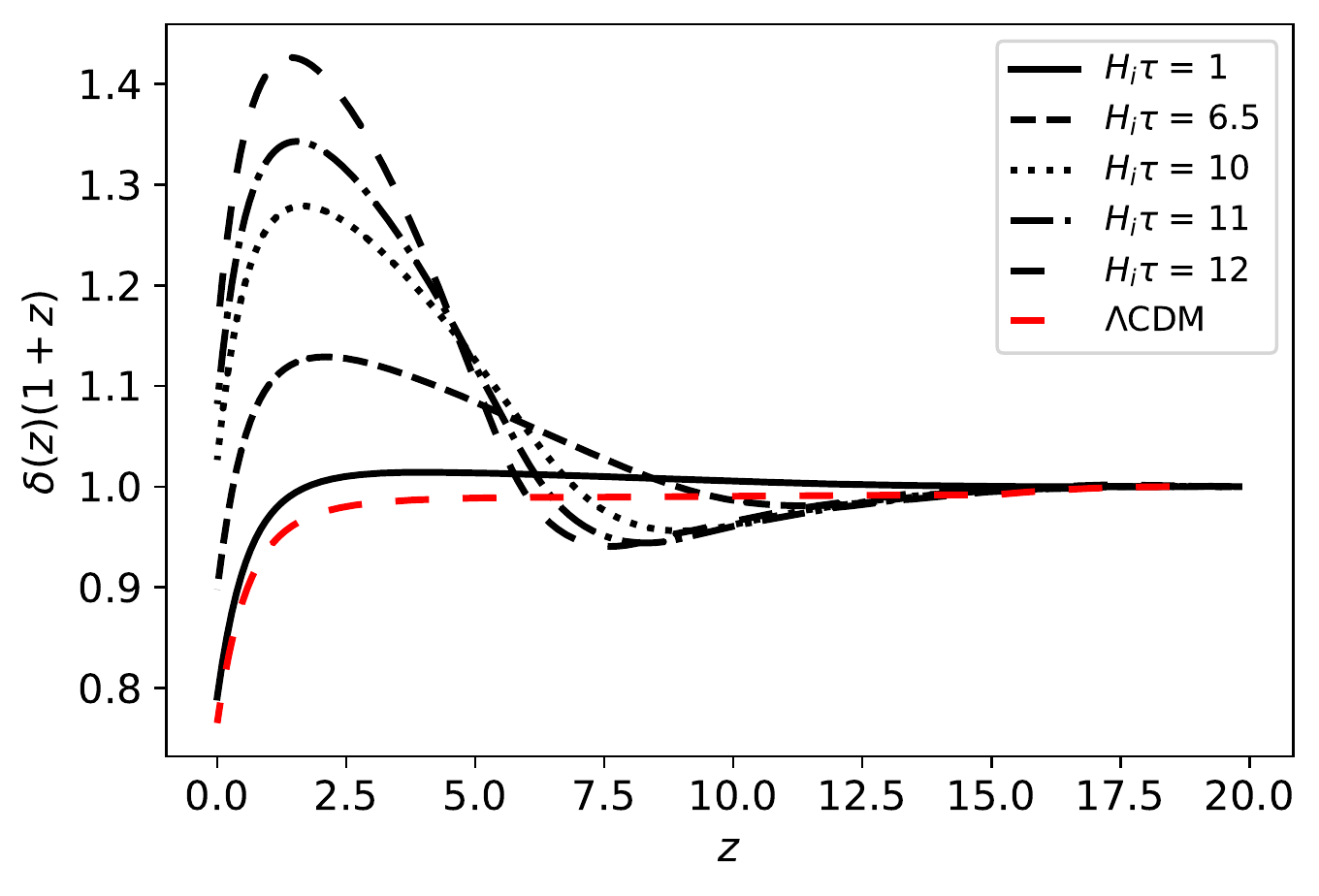}}}}
    \quad
    \subfloat[Ratio of the rate of growth to the rate of expansion\label{fig:dprime_over_aprime}]{\makebox[0.5\textwidth]{{\includegraphics[width=0.6\linewidth]{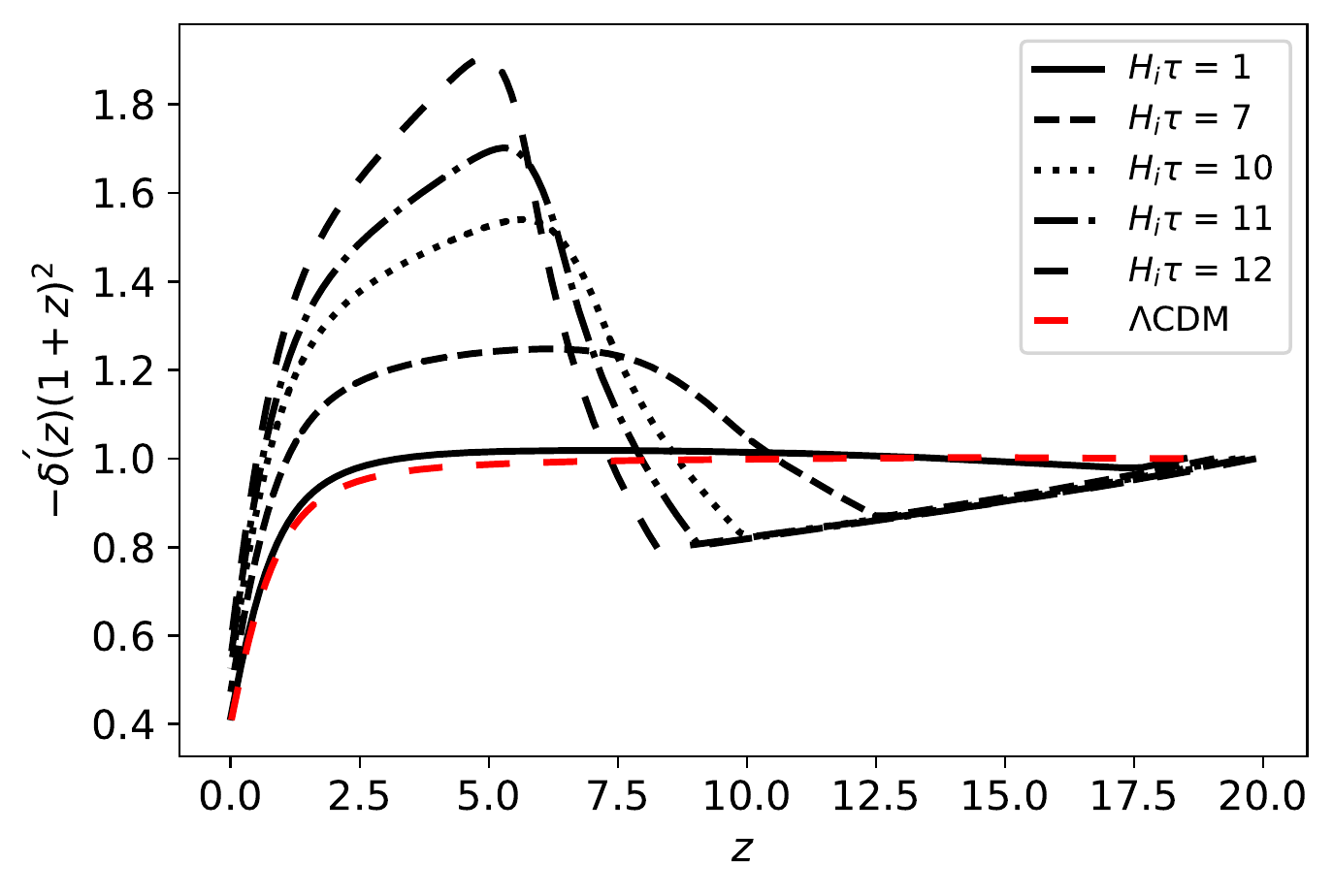}}}}
    \quad
    \caption{Comparison of the evolution of the density contrast and the scale factor as well as their time rates of change.}
\end{figure}

The characteristic decreasing of the growth rate predictions earlier on implies that, for a certain period, the delayed Universe was expanding faster than the perturbations were growing. We can see this clearly in Figure \ref{fig:delta_over_a}. In Figure \ref{fig:dprime_over_aprime}, we can also see that the rate of expansion is initially greater than the rate of growth of the fluctuation. Combined together, these two scenarios suppress the growth rate at early times. But after some time, the growth rate predictions start to increase after decreasing. Notice that the transition to this increasing phase is very abrupt. Since our definition of the growth rate also involves $\dot{a}(t)$, we expect to see kinks just as we saw in the background evolution. Interestingly, the growth rate becomes greater than unity at a certain point, implying that in the delayed Universe, perturbations will eventually grow faster than the Universe is expanding. This is also clear in Figures \ref{fig:delta_over_a} and \ref{fig:dprime_over_aprime}. Later on, however, dark energy starts to dominate and the growth rate is eventually driven down.

Figure \ref{fig:growth_rate_vartau_close} shows a closer look at the growth rate up to redshift $z \sim 3$. Here, we can see that the kinks in the growth rate evolution can also provide an upper bound. While the uncertainties of growth rate data at $z > 2.5$ are very large, it is safe to say that $H_i\tau \approx 18$ is already very unlikely to be viable. On the other hand, time delays with magnitude $H_i\tau \lesssim 10$ do appear viable. 

\subsection{$f\sigma_8(z)$}

\begin{figure}
    \centering
    \includegraphics[width=0.6\linewidth]{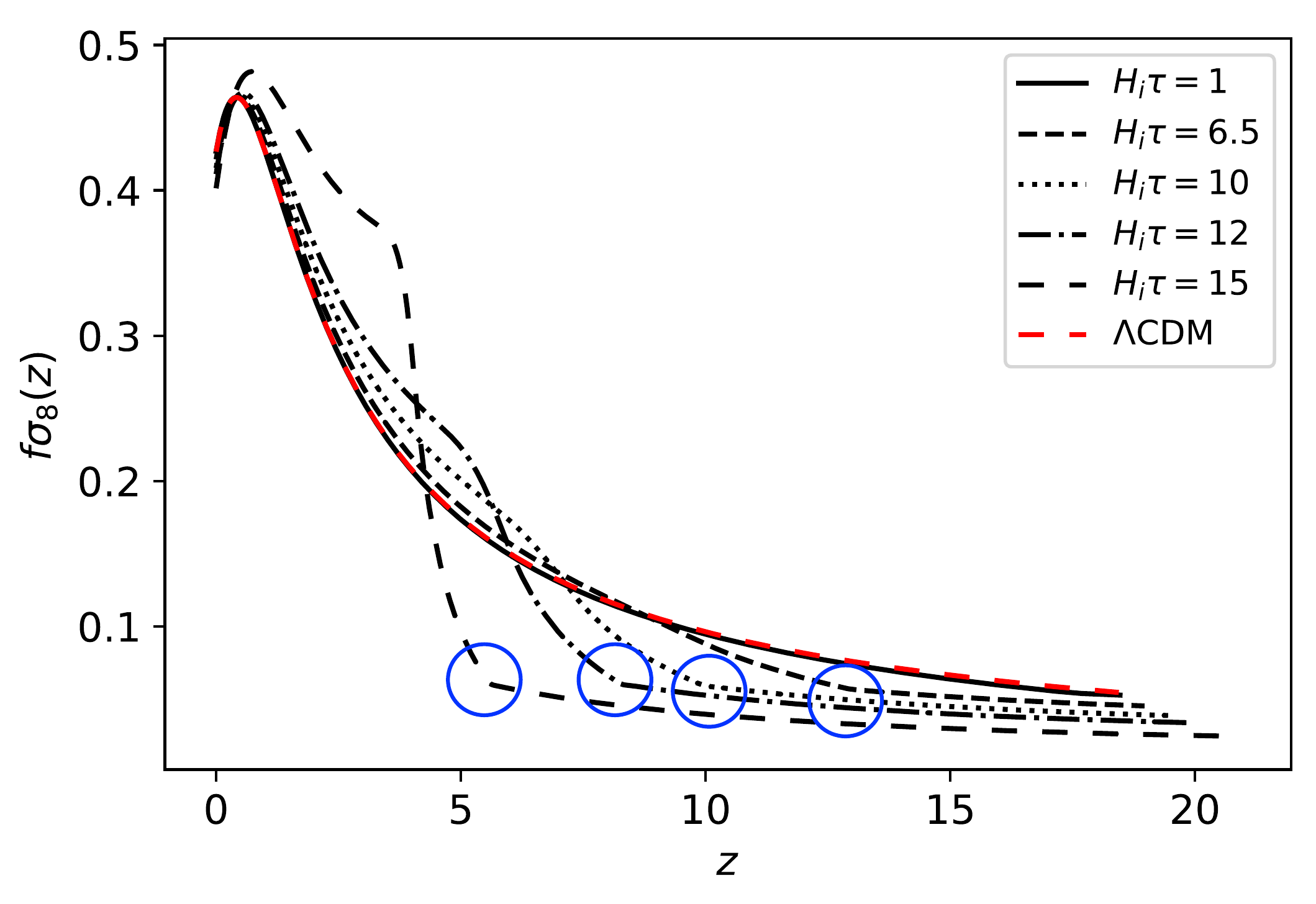}
    \caption{The evolution of $f\sigma_8(z)$. Time-delayed cosmology with intermediate (i.e. $H_i\tau \sim 10$) time delay parameter values predict markedly distinct $f\sigma_8(z)$ evolutions. It is not as pronounced here, but there are also kinks (encircled in blue) in this plot for the time-delayed cosmology predictions.}
    \label{fig:fsigma8_vartau_full}
\end{figure}

Figure \ref{fig:fsigma8_vartau_full} shows the plot of $f\sigma_8(z)$. Again, the standard $\Lambda$CDM prediction is shown in dashed red, and the black curves are the predictions of time-delayed cosmology at different values of the time delay parameter.  Similar to the scenario with the growth rate, time-delayed cosmology models with intermediate time delay parameter values predict $f\sigma_8(z)$ evolutions that are different from $\Lambda$CDM. The $f\sigma_8(z)$ of the delayed Universe starts off smaller than but eventually surpasses the standard prediction. When the cosmological constant becomes more important than matter during the dark energy-dominated era, time-delayed cosmology and $\Lambda$CDM follow each other in similar evolutions. Naturally, we also find that smaller delays lead to $f\sigma_8(z)$ predictions that are indistinguishable from $\Lambda$CDM. As with the growth rate, we also get a kink because the definition of $f\sigma_8(t)$ also includes $\dot{a}(t)$. Notice that in Figure \ref{fig:fsigma8_vartau_full} the predictions start out at different magnitudes. Note that all calculations started out with the same initial condition. The differences in the initial value in these plots are due to the normalizing constant $\delta(t = t_0)$, which is of course different for different models.

\begin{figure}
    \centering
    \subfloat{\makebox[0.5\textwidth]{{\includegraphics[width=0.6\linewidth]{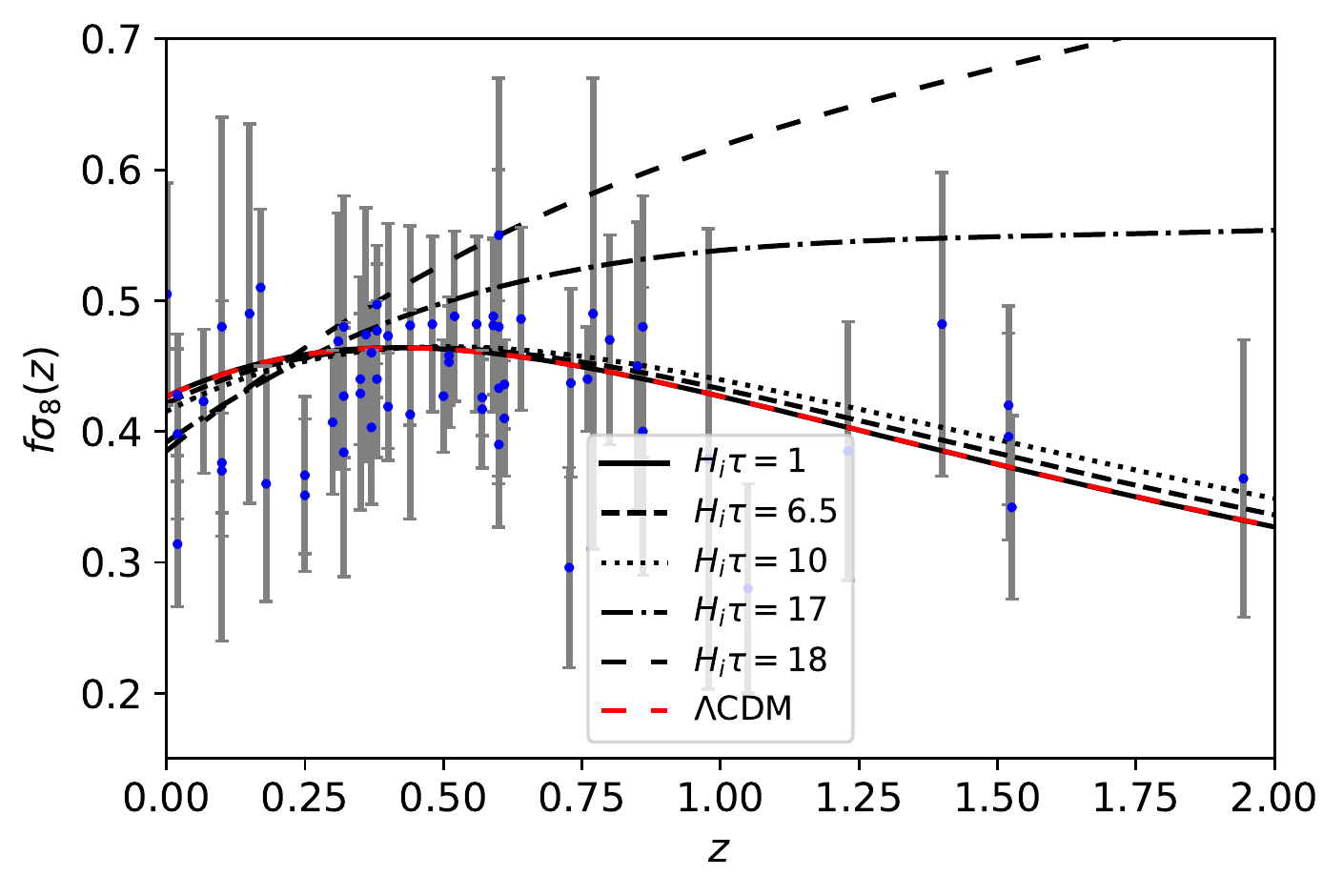}}}}
    \quad
    \subfloat{\makebox[0.5\textwidth]{{\includegraphics[width=0.6\linewidth]{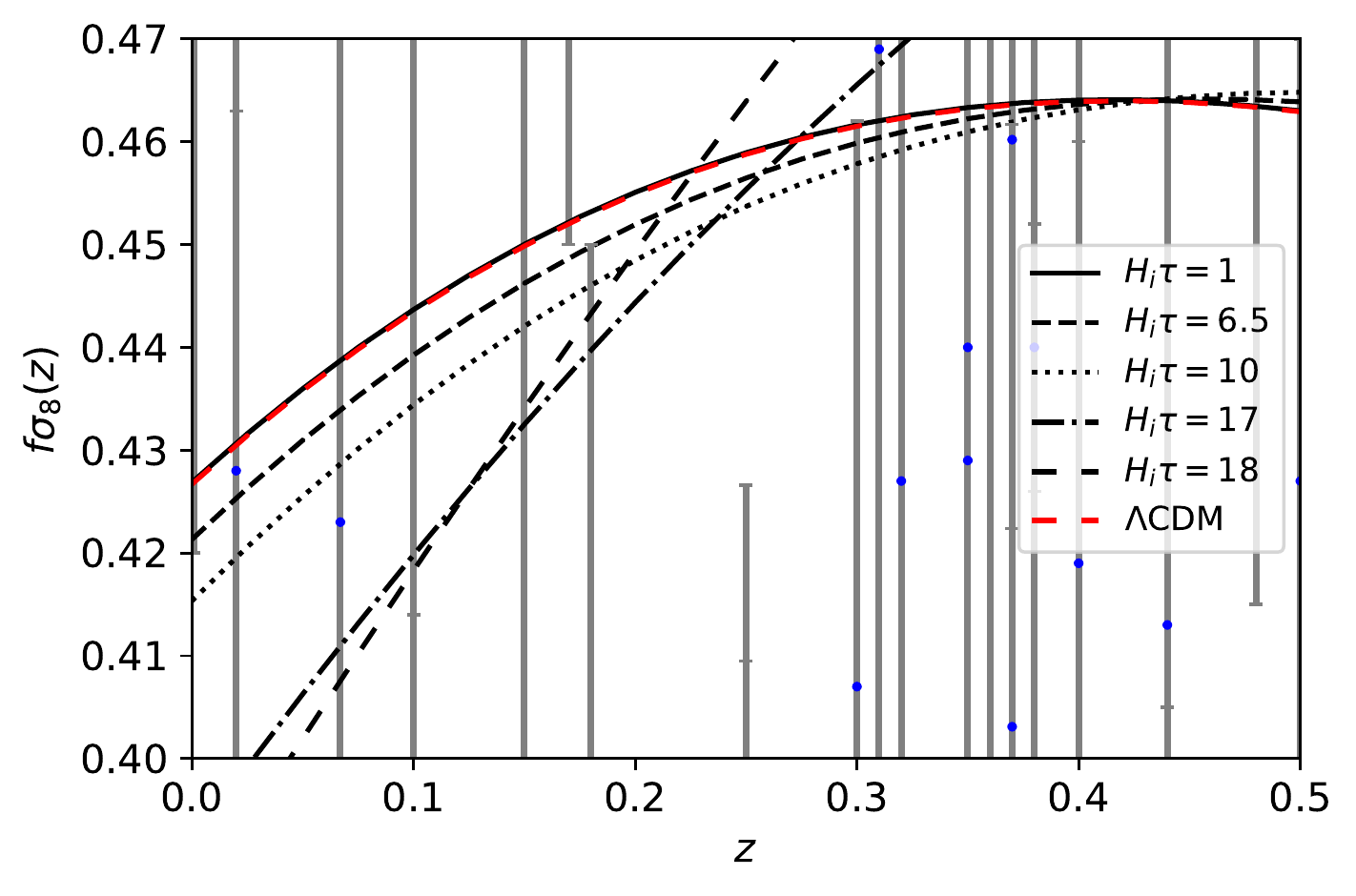}}}}
    \quad
    \caption{The evolution of $f\sigma_8(z)$ at small  and intermediate redshifts. The $f\sigma_8(z)$ data are taken from the compilation in Ref. \cite{Kazantzidis:2018rnb}. }\label{fig:fsigma8_vartau_close}
\end{figure}

Figure \ref{fig:fsigma8_vartau_close} is a closer look at $f\sigma_8(z)$ up to redshift $z = 2$. In this case, we do not see any kink even for a time delay $H_i\tau \approx 18$. However, it is clear that a time delay $H_i\tau \approx 18$ is already unlikely since its predicted evolution already misses plenty of data points. Time delays $H_i\tau \lesssim 10$ do however lead to predictions that are already distinct from $\Lambda$CDM while also appearing viable.

\section{Delay estimate}\label{estimate}

We have already obtained a strict upper bound for the time delay but to achieve a best estimate, we confront our numerical solutions for $H(z)$, $f(z)$, and $f\sigma_8(z)$ with observational data using a Markov chain Monte Carlo (MCMC) analysis. Given data $d$ and parameters $p$ of a model $m$, Bayes' theorem states that 
\begin{equation}
    P(p | d, m) = \dfrac{P(d | p, m)P(p | m)}{P(d|m)},
\end{equation}
where $P(p | d, m)$ (the posterior) is the probability distribution of the parameters $p$ given $d$ and $m$, $P(d | p, m)$ (the likelihood) is the probability of getting the data $d$ given $p$ and $m$, $P(p | m)$ (the prior) is the probability of the parameters $p$ according to our prior beliefs, and finally $P(d | m)$ (the evidence) is a normalizing constant that, as we shall see, is important for model comparison. 

\begin{table}[tbp]
\centering
\begin{tabular}{|c|c|}
\hline
Parameter & Prior \\
\hline
$H_i\tau$ & $[0,20]$\\
$H_0$ & $[20,100]$ \\
$\sigma_8$ & $[0.5,0.9]$\\
\hline
\end{tabular}
\caption{\label{tab:priors} We choose uniform priors defined over wide ranges for all the parameters.}
\end{table}

\begin{figure}
    \centering
    \subfloat{\makebox[0.5\textwidth]{{\includegraphics[width=0.6\linewidth]{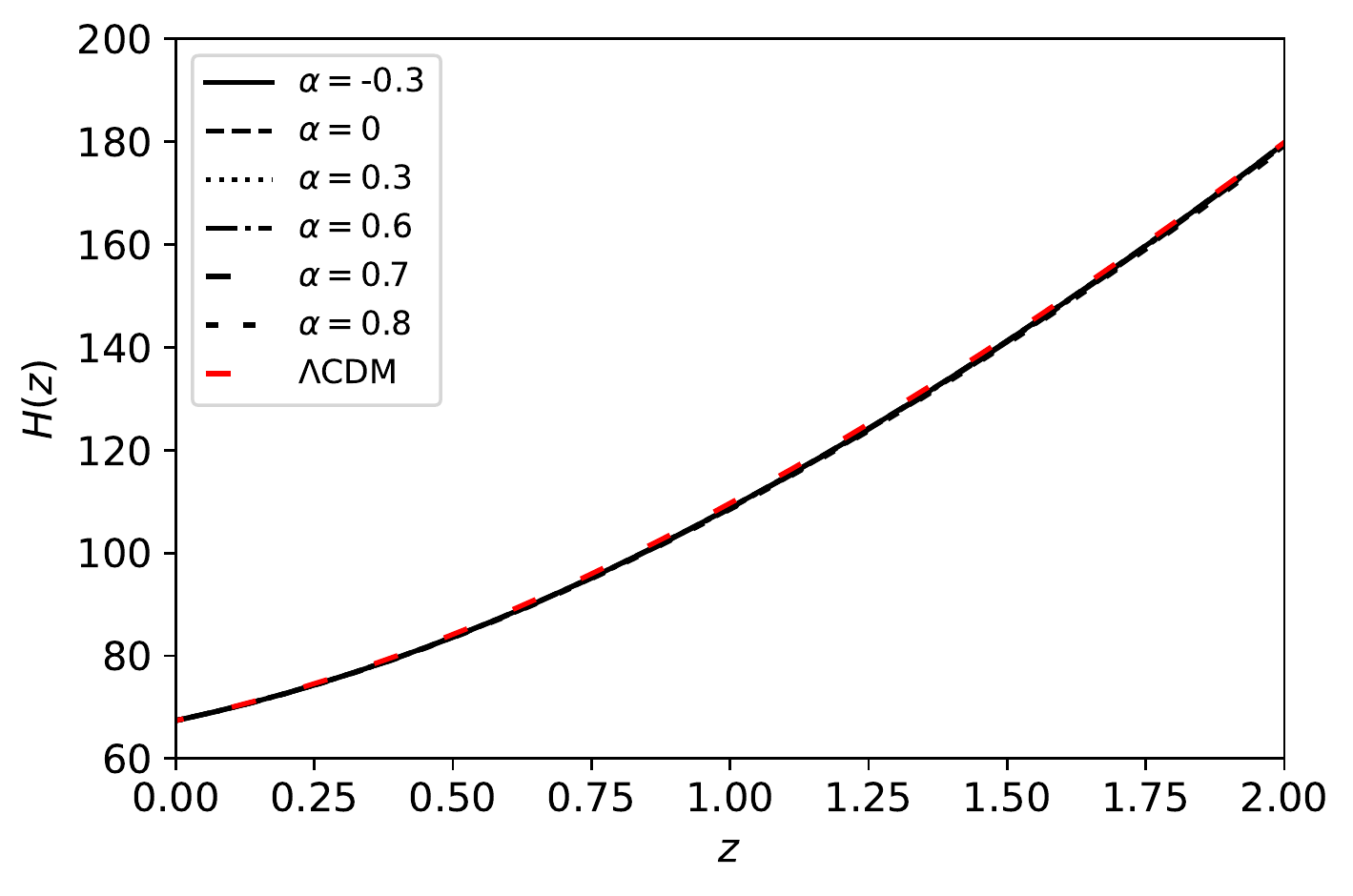}}}}
    \quad
    \subfloat{\makebox[0.5\textwidth]{{\includegraphics[width=0.6\linewidth]{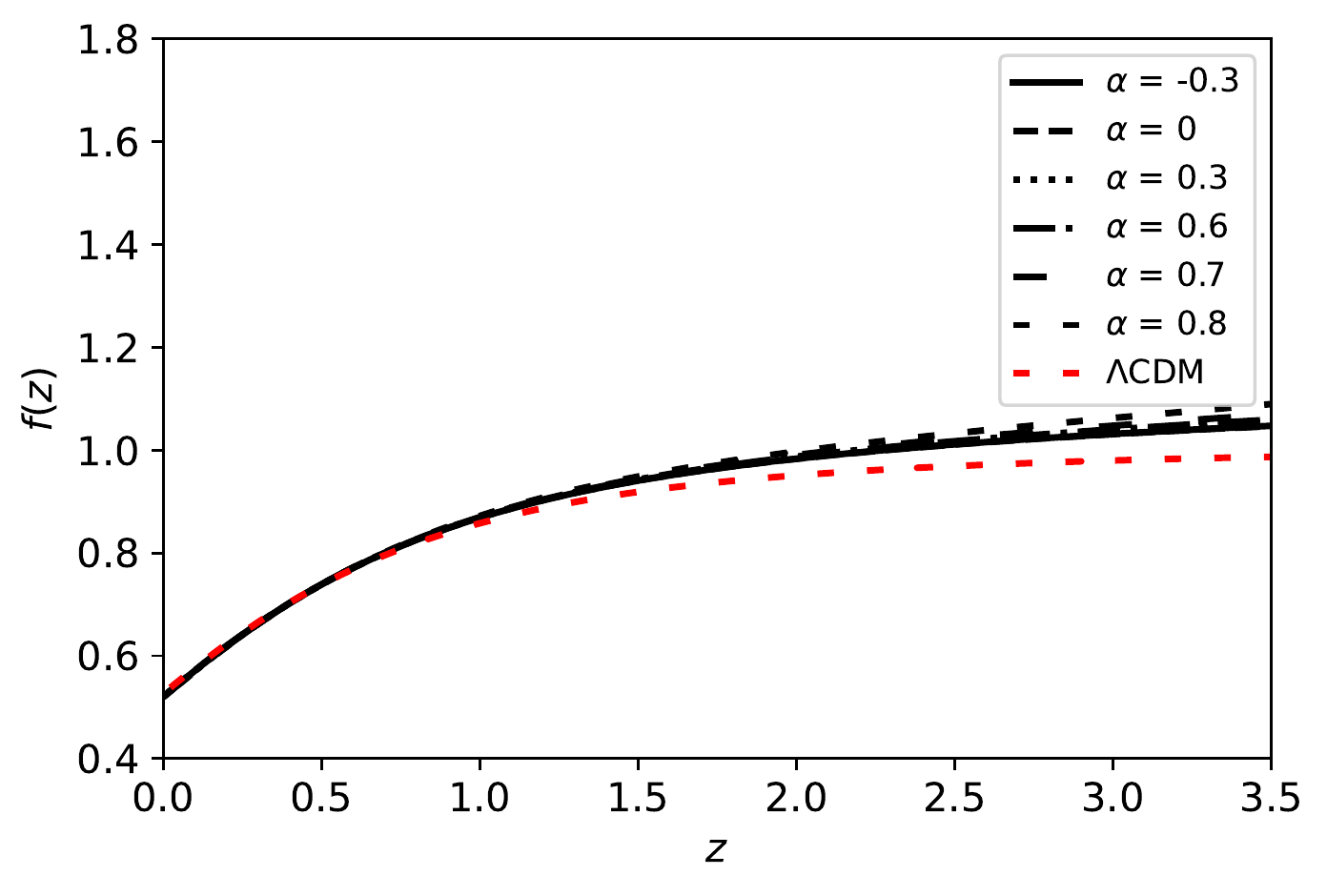}}}}
    \quad
    \subfloat{\makebox[0.5\textwidth]{{\includegraphics[width=0.6\linewidth]{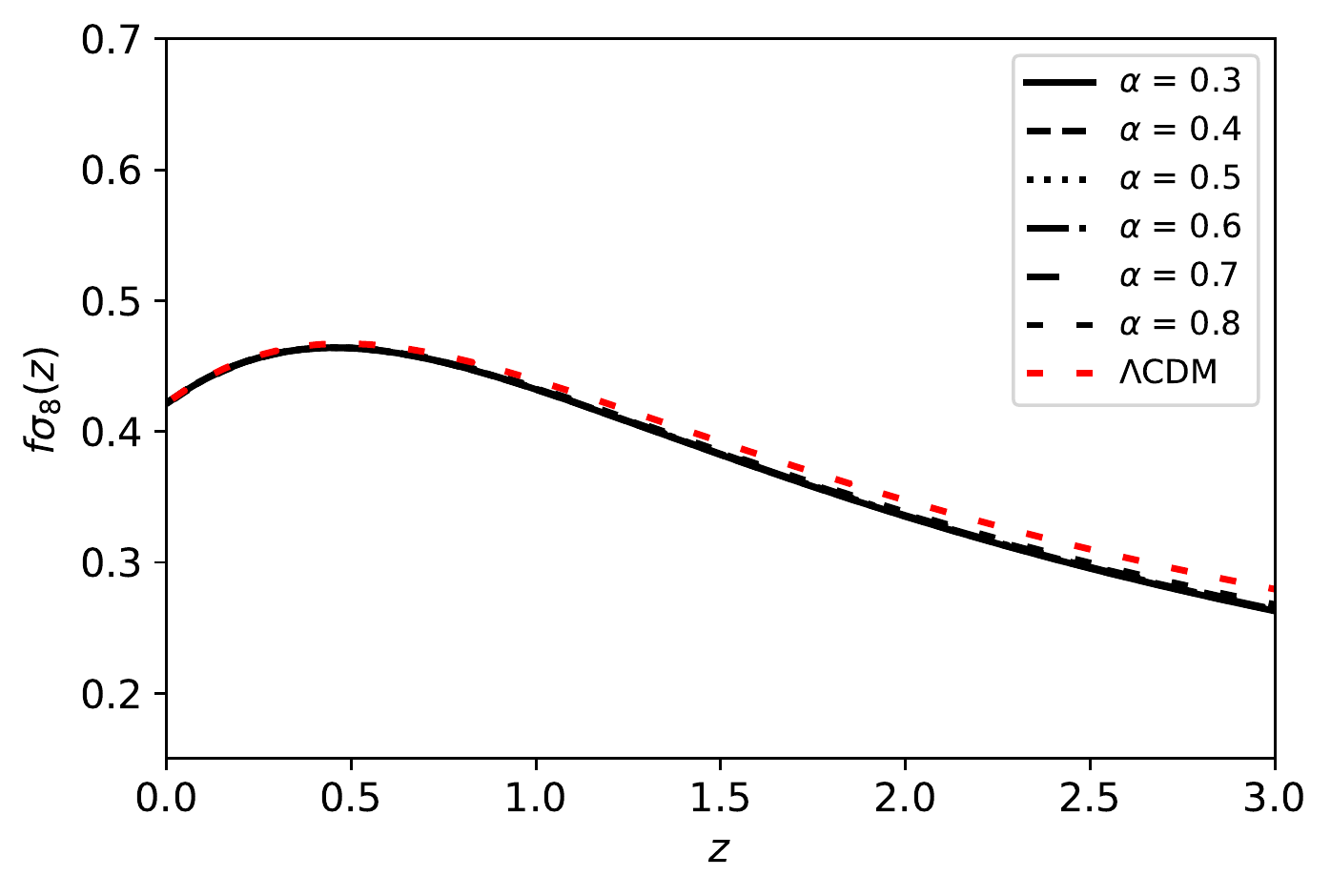}}}}
    \caption{The evolution of different observables for a fixed time delay and at varying initial histories. Here, $H_i\tau = 6.5$. The effect of $\alpha$ on observables is not strong at small redshifts and for time delays $H_i\tau < 10$.}\label{fig:all_varalpha_close}
\end{figure}

We consider a likelihood $L$ given by
\begin{equation}
    \ln L \sim -\sum_i^N\dfrac{(\mu^\text{obs}_i-\mu^\text{th}_i)^2}{2\sigma_i^2},
\end{equation}
where $N$ is the number of data points, $\mu^\text{obs}_i$ is an observational data point, $\mu^\text{th}_i$ is a predicted value, and $\sigma_i$ is the observational error. Aside from the time delay, we also take the Hubble constant $H_0$ and $\sigma_8$ as free parameters to be estimated. Again, we fix $\alpha$ to 2/3 because $\alpha$ has a weak effect on the observables at small redshifts (see Figure \ref{fig:all_varalpha_close}) and because this value of $\alpha$ gives the canonical matter-era solution. 

Our priors are shown in Table \ref{tab:priors}. We intentionally choose priors defined over wide ranges so as to avoid inadvertently cutting the posterior short. We note that since our priors are uniform, our arbitrary cutoffs for the priors do not affect the value of the best estimates of the parameters so long as the priors include these best estimates in their ranges. Since each of our prior is defined over a wide range, the best estimate for a parameter is guaranteed to be within the prior for that parameter. 

Note that we do not consider negative delays (i.e. $H_i\tau < 0$). A negative delay means that the delayed Friedmann equation is advanced in time rather than retarded. In which case, we must provide future information instead of an initial history. The solution then would be the past evolution. Since we are interested in the predictions of time-delayed cosmology in the late Universe, the time delay must be strictly positive.

\begin{figure}
    \centering
    \includegraphics[width=0.6\linewidth]{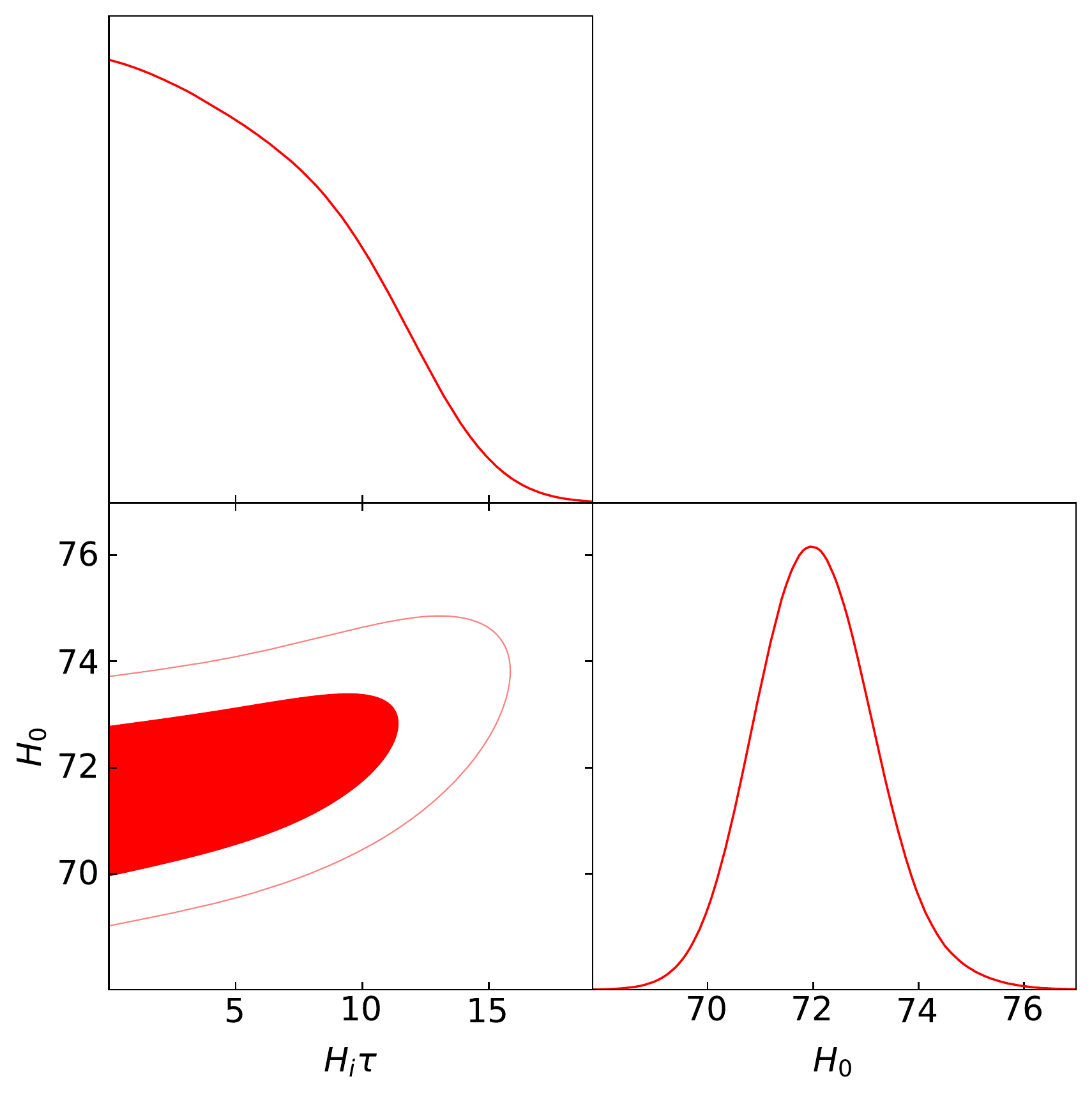}
    \caption{Posterior distributions of the time delay parameter $\tau$ and the Hubble constant $H_0$ using Hubble data. The median estimate is $H_i\tau =  5.59^{+4.89}_{-3.86}$ or $\tau = 0.098^{+0.085}_{-0.068}$ Gyr, while the median estimate for the Hubble constant is $H_0 = 72.02^{+1.13}_{-1.07}$ km s$^{-1}$ Mpc$^{-1}$. }
    \label{fig:hubble_pos}
\end{figure}

We use the \verb|PyMultiNest| \cite{Buchner:2014nha} and \verb|GetDist| \cite{Lewis:2019xzd} Python packages to sample the posteriors via MCMC and post-process the resulting MCMC chains. We consider the Hubble expansion rate data compiled in Ref. \cite{Gogoi:2021mhi}, the growth rate data compiled in Refs. \cite{Mirzatuny:2013nqa} and \cite{Peel:2012vg}, and $f\sigma_8(z)$ data compiled in Ref. \cite{Kazantzidis:2018rnb}. In what follows, we choose to report the median estimate which is more robust to outliers as compared with the mean. We have checked however that the median estimates below are not too different from the mean estimates and overlap with them within $1\sigma$.

Figure \ref{fig:hubble_pos} shows the posterior distributions for the time delay parameter $\tau$ and the Hubble constant $H_0$ using Hubble expansion rate data alone. The median estimate for the time delay is $H_i\tau =  5.59^{+4.89}_{-3.86}$ or $\tau = 0.098^{+0.085}_{-0.068}$ Gyr. Meanwhile, the median estimate for the Hubble constant is $H_0 = 72.02^{+1.13}_{-1.07}$ km s$^{-1}$ Mpc$^{-1}$. Notably, the credible interval of the time delay estimate is rather large and this is the case for all the data we consider in this work. This may be attributed to two things. Firstly, the uncertainties of the observational data points are themselves large. And secondly, from Figure \ref{fig:hubble_vartau_close}, we can't expect a sharply peaked posterior with a narrow credible interval because the predictions of time-delayed cosmology for varying time-delays are very similar. 

\begin{figure}
    \centering
    \includegraphics[width=0.35\linewidth]{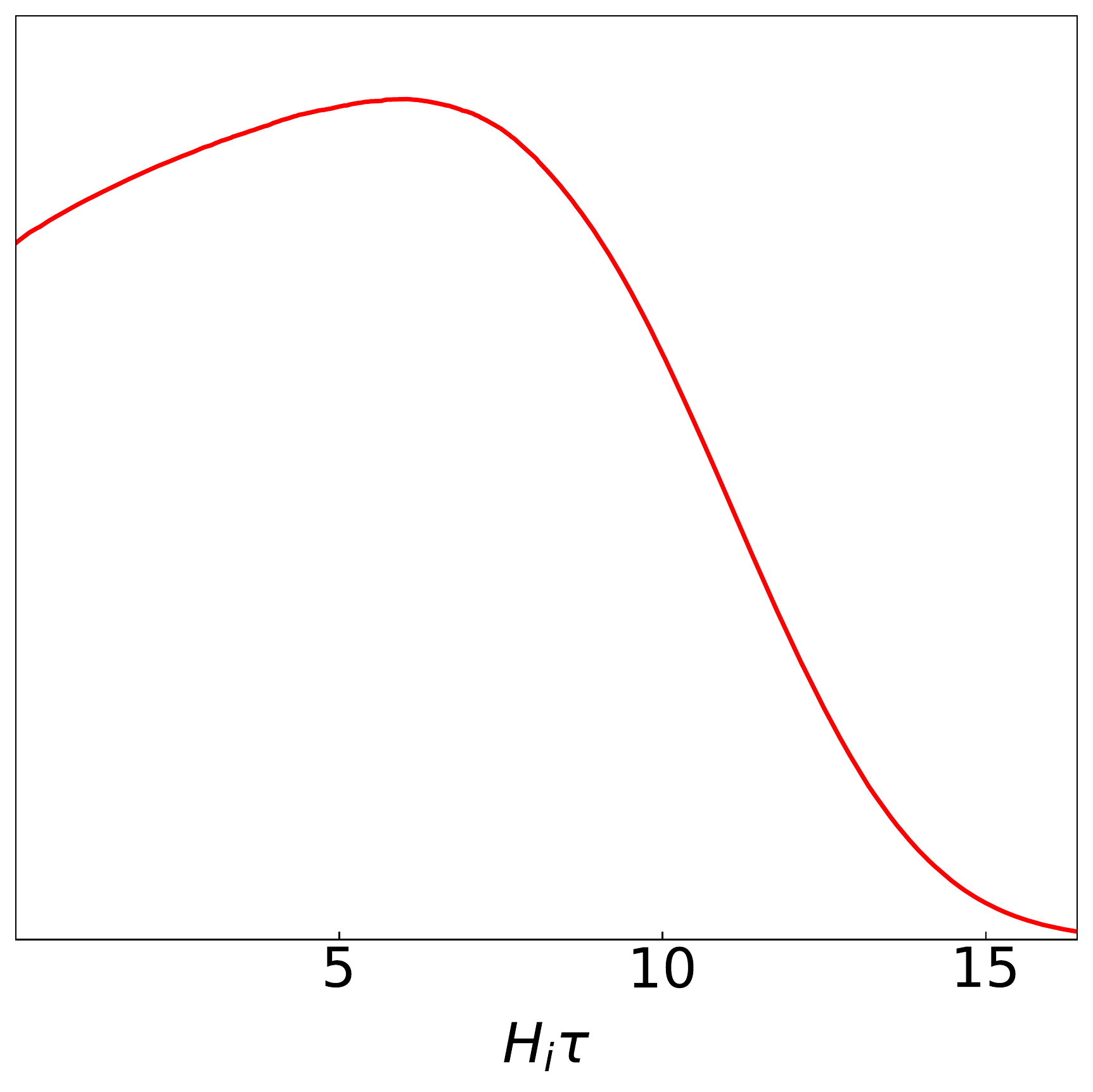}
    \caption{Posterior distribution of the time delay parameter $\tau$ using the growth rate data set alone.   The median estimate is $H_i\tau =  6.10^{+4.04}_{-4.10}$ or $\tau = 0.106^{+0.072}_{-0.072}$ Gyr.}
    \label{fig:f_pos}
\end{figure}

\begin{figure}
    \centering
    \includegraphics[width=0.6\linewidth]{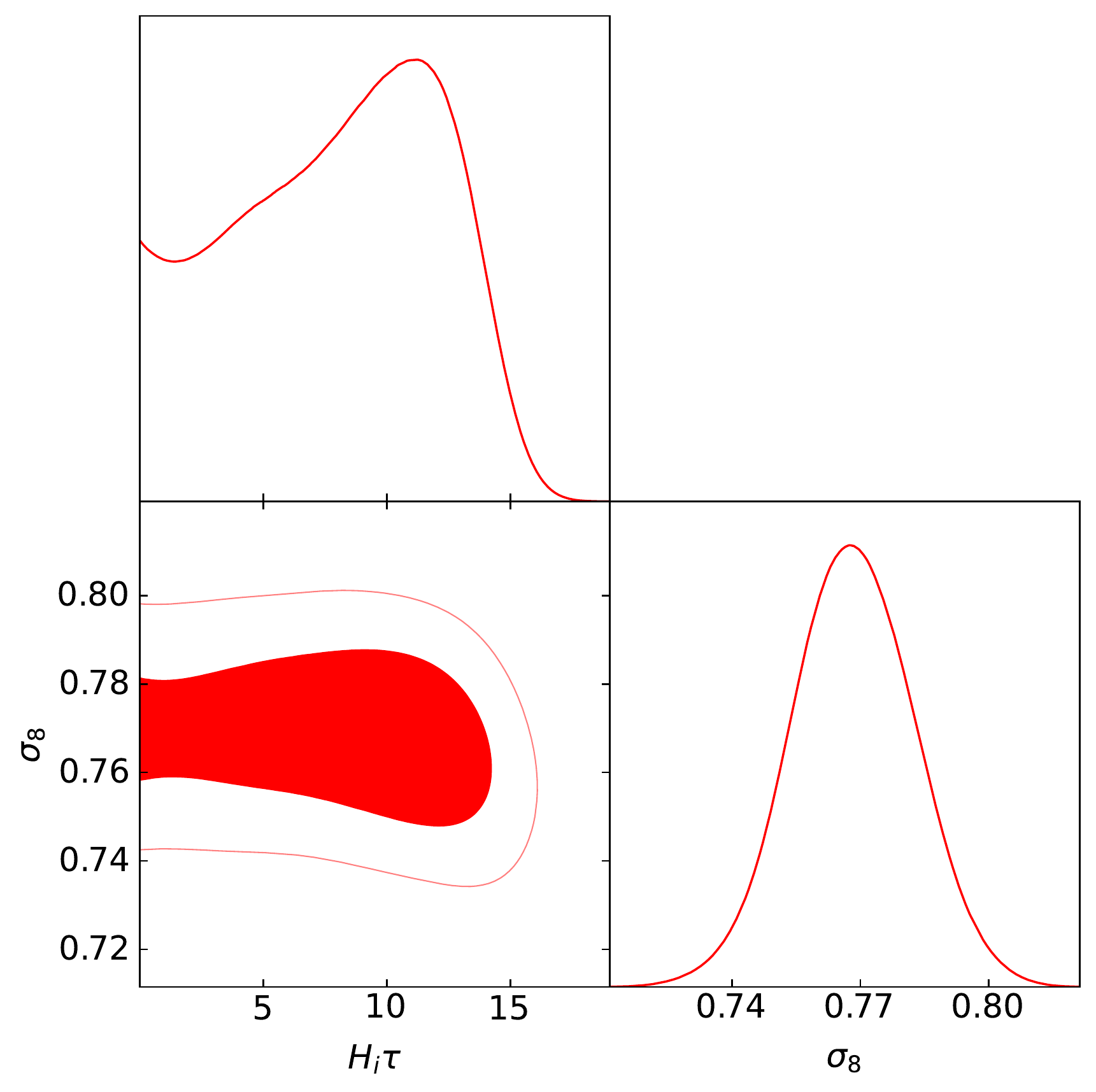}
    \caption{Posterior distributions of the time delay parameter $\tau$ and $\sigma_8$ using the $f\sigma_8(z)$ data alone. The median estimates are $H_i\tau = 8.58^{+4.04}_{-5.30}$ or $\tau = 0.150^{+0.068}_{-0.091}$ Gyr and $\sigma_8 = 0.77 \pm 0.01$.}
    \label{fig:fsigma8_pos}
\end{figure}

\begin{figure}
    \centering
    \includegraphics[width=0.6\linewidth]{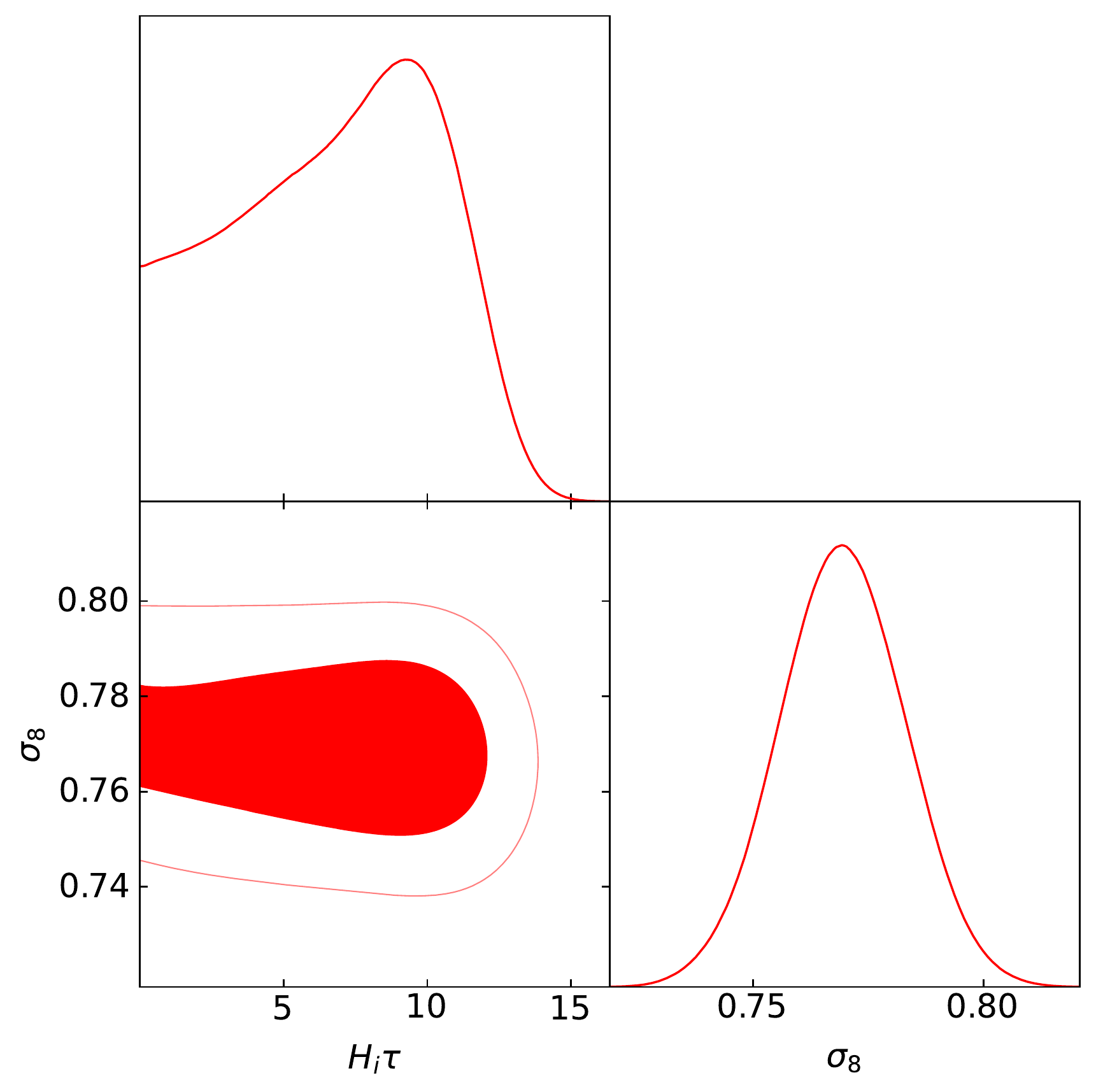}
    \caption{Posterior distribution of the time delay parameter $\tau$ and $\sigma_8$ using the combined growth rate and $f\sigma_8(z)$ datasets. The median estimates are $H_i\tau = 7.26^{+3.35}_{-4.57}$ or $\tau = 0.125^{+0.060}_{-0.080}$ Gyr and $\sigma_8 = 0.77 \pm 0.01$.}
    \label{fig:perturbation_pos}
\end{figure}

Figure \ref{fig:f_pos} shows the posterior distribution for the time delay parameter $\tau$ using the growth rate dataset alone. Upon sampling, we find that the median estimate for the time delay is $H_i\tau =  6.10^{+4.04}_{-4.10}$ or $\tau = 0.106^{+0.072}_{-0.072}$ Gyr. The estimate for $\tau$ has notably increased and we also find that the mass of the posterior distribution has moved to a nonzero time delay. On the other hand, Figure \ref{fig:fsigma8_pos} shows the posterior distributions for the time delay parameter $\tau$ and $\sigma_8$ using the $f\sigma_8(z)$ dataset. The median estimate for the time delay is $H_i\tau = 8.58^{+4.04}_{-5.30}$ or $\tau = 0.150^{+0.068}_{-0.091}$ Gyr. The median estimate for $\sigma_8$ is $0.77 \pm 0.01$.  Notice in this case that the estimate for the time delay has gotten much larger. Figure \ref{fig:perturbation_pos} shows the posterior distributions when we combine the growth rate and $f\sigma_8(z)$ datasets.  We find that the median estimates for the parameters are $H_i\tau = 7.26^{+3.35}_{-4.57}$ or $\tau = 0.125^{+0.060}_{-0.080}$ Gyr and $\sigma_8 = 0.77 \pm 0.01$. What these results show is that growth observables or perturbations consistently prefer nonzero values of the time delay parameter, especially $f\sigma_8$ data. We can see this not only in the median estimate but also in the mode.

\begin{figure}
    \centering
    \includegraphics[width=0.8\linewidth]{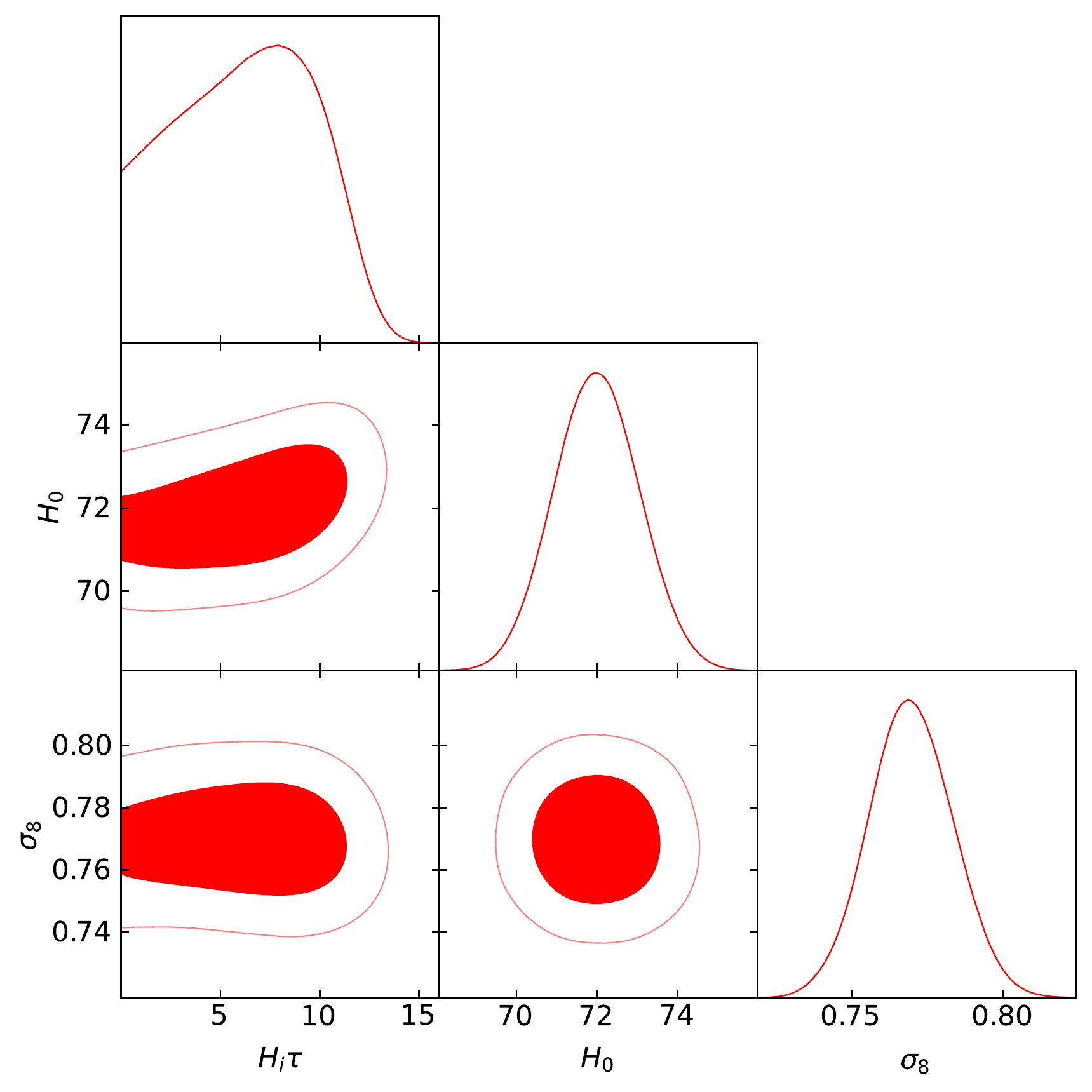}
    \caption{Posterior distributions of the time delay parameter $\tau$, the Hubble constant $H_0$, and $\sigma_8$ using the combined background and growth datasets. The median estimates are $H_i\tau = 6.49^{+3.52}_{-4.01}$ or $\tau = 0.113^{+0.060}_{-0.069}$ Gyr, $H_0 = 72.00^{+1.05}_{-1.04}$ km s$^{-1}$ Mpc$^{-1}$, and $\sigma_8 = 0.77 \pm 0.01$.}
    \label{fig:all_pos}
\end{figure}

To arrive at a best estimate, we combine the background and growth datasets. Figure \ref{fig:all_pos} shows the posterior distributions of the time delay parameter, the Hubble constant $H_0$, and $\sigma_8$. The median estimates are $H_i\tau = 6.49^{+3.52}_{-4.01}$ or $\tau = 0.113^{+0.060}_{-0.069}$ Gyr, $H_0 = 72.00^{+1.05}_{-1.04}$ km s$^{-1}$ Mpc$^{-1}$, and $\sigma_8 = 0.77 \pm 0.01$. The best estimate for the time delay is expectedly between the background median estimate and growth median estimate. It is clear from the results that growth observables especially prefer higher values of the time delay. 

To further strengthen our statistical analysis of time-delayed cosmology, we compute the Bayes factor which is roughly the Bayesian equivalent of the $p-$value used for classical (frequentist) hypothesis testing. Given data $d$ and two models $m_1$ and $m_2$, the preference for $m_1$ over $m_2$ in light of $d$ is quantified by the Bayes factor $B_{12}$ defined as
\begin{equation}
    B_{12} := \dfrac{P(d|m_1)}{P(d|m_2)},
\end{equation}
which is simply the ratio of the evidence of $m_1$ to the evidence of $m_2$. This definition assumes that both models are equally probable \textit{before} accounting for the data. This is a fair assumption in our case since this is the first time that a time delay is even being considered in late-time cosmology and we do not have prior information whether time-delayed cosmology is preferred over $\Lambda$CDM.

Letting $m_1$ denote $\Lambda$CDM and $m_2$ denote time-delayed cosmology, we compute the Bayes factor based on evidences calculated using the Hubble expansion rate data alone ($\ln B_{12} = 0.446 \pm 0.111$), the combined growth data ($\ln B_{12} = 0.100 \pm 0.092$), and finally the combined background and growth data ($\ln B_{12} = 0.175 \pm 0.142$). Following the criteria in Ref. \cite{Trotta:2008qt}, we find that regardless of the data considered, the Bayes factor indicates a statistical preference against time-delayed cosmology in favor of $\Lambda$CDM that is not worth more than a bare mention (an odds in favor of $\Lambda$CDM \textit{less} than 3:1). In other words, no conclusion can be drawn as to which model is favored.

\section{Conclusion}\label{conclusion}

This paper has made initial steps towards confronting the predictions of time-delayed cosmology with data. We have applied the delayed Friedmann equation in the late-time Universe and chose the Hubble expansion rate $H(z)$ and Newtonian matter perturbations as our observational probes. We obtained the predictions for the late-time background evolution and the growth data. In calculating the growth observables, we have used the standard perturbation equation and assumed that the effects of time-delayed cosmology enter through the background expansion only. We find that the conservative assumptions we have made are sufficient to reveal smoking-gun imprints of the phenomenological time delay. These imprints can be credited to the propagation of discontinuities inherent in the solutions of delay differential equations. This is the first time that the effects of these discontinuities have been demonstrated in this model.

We showed that for ``intermediate'' (i.e. $\tau \sim 0.175 \pm 0.001$ Gyr) values of the time delay parameter, time-delayed cosmology already makes different predictions as compared to $\Lambda$CDM, especially when looking at growth observables. The difference can be seen at redshifts that are currently accessible to us. Our best estimate of the key time delay parameter is $\tau = 0.113^{+0.060}_{-0.069}$ Gyr using the combined Hubble expansion rate and growth datasets. Our calculation shows that the key time delay parameter does not have to be in orders of Planck time as originally proposed in order to be consistent with observations. We also calculated the Bayes factor and find no conclusive evidence in favor of $\Lambda$CDM against time-delayed cosmology. To our knowledge, this study is the first systematic attempt to place a statistical and data-driven constraint on time-delayed cosmology as applied to late times. 

While we are mainly interested in the late Universe, one can take our calculated value of the time delay and ask what it means for an inflationary time-delayed cosmology. The biggest implication of a time delay as large as our estimated value is that inflation will last for millions of years. This goes against the usual estimate of the period of inflation lasting for a tiny fraction of a second which is based on certain assumptions on initial conditions, e.g. inflation started at around $10^{-36}$ s after the initial singularity with a large Hubble parameter. If we relax these assumptions and allow a pre-Big Bang or ekpyrotic scenario, then it is not immediately evident how such a long period of inflation would be troublesome. It is the number of e-folds, and not the length, of inflation that is important. 

 Our results show that time-delayed cosmology is not only interesting theoretically, but it can also hold up against the standard $\Lambda$CDM model when confronted with currently available background and growth data. Our work provides a data-driven motivation to further study this phenomenological model. Future large-scale structure surveys \cite{Fanizza:2021tiv, Fanizza:2021tuh} and high redshift distance indicators such as proposed standardizable candles (quasars \cite{Bargiacchi:2021hdp} and gamma ray bursts \cite{DeSimone:2021rus}) and standard sirens \cite{Virgo:2021bbr, Palmese:2021mjm} can be expected to further constrain the time delay, if not rule it out completely should the kink inherent to a time-delayed solution not be observed. We leave the search for a fundamental action that leads to a time-delayed cosmology for future work.

\begin{acknowledgments}\label{sec:acknowledgements}
The authors thank Che-Yu Chen, Kin-Wang Ng, and Jackson Levi Said for helpful comments on an earlier version of the manuscript. RCB is supported by the Institute of Physics, Academia Sinica. This
research is supported by the University of the Philippines Diliman Office of the Vice Chancellor for Research and Development through Project No. 191937 ORG.
\end{acknowledgments}

\bibliographystyle{JHEP}
\bibliography{refs}

\end{document}